%% file: ms.tex
\documentclass[12pt,preprint]{aastex}

\slugcomment{To appear in the Astronomical Journal}

\shorttitle{3D Morphology of VY CMa, Kinematics}
\shortauthors{Humphreys  et al.}

\begin{document}


\title{The 3D Morphology of VY Canis Majoris. I The Kinematics of the Ejecta\footnote{Based on observations with the NASA/ESA
{\it Hubble Space Telescope} obtained at the Space Telescop Science Institute, which is 
operated by the Association of Universities for Research in Astronomy, Inc. under NASA
contract NAS 5-26555.}}


\author{Roberta M. Humphreys, L. Andrew Helton and Terry J. Jones}
\affil{Astronomy Department, University of Minnesota, Minneapolis, MN 55455 }


\begin{abstract}
Images of the complex circumstellar nebula associated with the famous red supergiant 
VY CMa show evidence
for multiple and asymmetric mass loss events over the past 1000 yrs. Doppler velocities of 
the arcs and knots in the ejecta showed that they are not only spatially distinct but also
kinematically separate from the surrounding diffuse material. In this paper we describe 
second epoch HST/WFPC2 images to measure the transverse motions which when combined with the
radial motions provide a complete picture of the kinematics of the ejecta including the
total space  motions and directions of the outflows.  Our results show that
the arcs and clumps of knots are moving at different velocities, in different directions,
and at different angles relative to the plane of the sky and to the star, confirming 
their origin from eruptions at different times and from physically  separate regions on the star. 
We conclude that the morphology and kinematics of the arcs and knots are consistent with a
history of mass ejections not aligned with any presumed axis of symmetry.
The arcs and clumps represent relatively massive outflows and ejections of gas very likely
associated with large -- scale convective activity and magnetic fields.

\end{abstract}


\keywords{circumstellar matter --- supergiants ---  stars:winds, outflows --- stars:activity --- stars:individual(VY CMa) }

\section{Introduction}

The extreme red supergiant and powerful infrared source and OH maser, \object{VY Canis Majoris}
is one of the most luminous evolved stars known. At its distance of 1.5 kpc (Herbig 1972; Lada \& Reid 1978; Marvel 1997), VY CMa's luminosity is $\approx$ 4.3 $\times 10^{5} L_{\odot}$ ( Appendix A).  Its very visible asymmetric nebula, 10$\arcsec$ across, combined with its high mass loss rate 
of $4 \times  10^{-4} M_{\odot} yr^{-1}$ \citep{Dan94}, makes VY CMa a special case even among the cool
hypergiants that define the upper luminosity boundary in the HR Diagram \citep{HD94,deJ98}.
VY CMa is ejecting large amounts of gas and dust at a prodigious rate, and is consequently one of 
our most
important stars for understanding the high mass loss episodes near the end of massive star
evolution.

High resolution imaging with HST (FOC, Kastner \& Weintraub 1998; WFPC2, Smith et al 2001)
and near--IR interferometry (Monnier et al 1999)  have revealed VY CMa's complex circumstellar environment.
The multi-wavelength HST/WFPC2 images of VY CMa (Smith et al ) especially showed 
the  complexty of detail in its ejecta including  the prominent nebulous arc to the
northwest, which is also visible in groundbased data,  two bright filamentary arcs to the 
southwest, plus
relatively bright clumps of dusty knots near the star, and numerous small arcs 
throughout the nebula. All of which are evidence for 
multiple and asymmetric mass loss episodes. The apparent random orientations 
of the arcs suggested that they were produced  by localized ejections, not necessarily aligned with 
either the star's presumed NE/SW axis (Morris \& Bowers 1980; Bowers et al 1983, Richards et al 1998)  or its equator. 
Smith et al therefore  speculated that the arcs may be expanding loops caused by localized activity on the star's
ill-defined surface. 

To learn more about the morphology, kinematics and origin of VY CMa's complex ejecta,
Humphreys et al (2005)  obtained long-slit spectra with HIRES on the Keck 1 telescope to map the 
emission and absorption lines in the nebula. The four slits were placed across several
structures in the nebula including the NW arc, the two outer filamentary arcs and  
clumps of bright knots. The Doppler motions of the reflected absorption lines and extremely strong
K I emission line revealed a complex pattern of velocities in the ejecta. Smith (2004) has also described spectra of the K I emission obtained later  at similar slit positions in the nebula, but 
published  velocities  for only a few positions. 
Humphreys et al found a strong velocity gradient across the NW arc which is expanding at 
$\sim$ 50 km s$^{-1}$ with respect to 
the embedded star, and  is ``kinematically distinct'' from the surrounding 
nebulosity. It was apparently ejected $\approx$ 400 yrs ago while the two outer filamentary
arcs were ejected in separate events possibly from 800 to 1000 yrs ago. Small arcs and knots closer to
the star were ejected more recently. Somewhat surprisedly, the more diffuse uniformly 
distributed gas and dust appeared to be essentially stationary with little or no radial velocity 
relative to the star.

Thus VY CMa  shows  evidence not only for multiple and asymmetric mass
loss events at different times, but also ejections that are recognizably and 
kinematically separate from the general flow of the diffuse material.  Obviously,
these results have serious implications for the origin of high mass loss events during 
the final stages of massive star evolution possibly  
involving convection and activity analogous to that in lower mass stars. However, the 
 overall expansion of the nebula, the direction of the outflows and orientation of the arcs within the surrounding 
nebulosity are not known. Fortunately,  
VY CMa provides us with a unique opportunity to determine the three dimensional  morphology
of its ejecta  and the geometry of the discrete structures embedded in it.  

Like other reflection nebulae, VY CMa is highly polarized. Herbig (1972) measured polarization
up to 70\% in the nebula, 
but no polarimetry has been done on VY CMa since his groundbased photographic measurements. 
We have obtained polarimetric images with the Advanced Camera for Surveys High Resolution 
Camera (ACS/HRC) on the HST together with second epoch WFPC2 images to measure the transverse 
motions.  The polarization together with the color from the images can yield information 
on the line of sight distribution of the nebulosity and let us determine the relative locations of the arcs and knots, while  the combined radial and transverse motions 
will provide the total 
 motion and direction of the outflow of the different features.

In this first paper we describe the second epoch images, our procedure for measuring 
the transverse motions of the arcs and knots (\S 2), and the resulting kinematics of the ejecta
(\S 3). In \S 4 we describe the geometry of the ejecta and  in the last section we discuss 
the high mass loss events from active regions on the 
star and the presence of magnetic fields in VY CMa's  ejecta. In the second paper we will present 
the polarimetry measurements and the resulting three dimensional morphology of this 
famous object.

\section{The HST Observations, Data Processing and Measurement Procedure}

The second epoch images of VY CMa were observed on  June 13, 2005 with the Planetary 
Camera on the WFPC2 with a pixel scale of 0$\farcs$0455. The  images were obtained 
using the medium width F410M, F547M and F1042M continuum filters and the 
F656N narrow band H-alpha filter.  The observing program, integration times etc.  
were identical to that described in Smith et al.  The polarimetric images were
made with the ACS/HRC on August 17, 2004 using the three visual polarizers 
(POL0V, POL60V, and POL120V) in two colors. The F550M(V) and F658N(red) filters 
 were chosen  because they are closest to those used for the WFPC2 images. A wide range of 
exposure times  allowed us to work close to the star and to image the fainter nebulosity.
All of the new images of VY CMa obtained for this program are summarized in 
Table 1.

To assure consistency of the data and our proper motion measurements we reduced
the 2005 images in tandem with the reprocessing of the first epoch images from March 22,
1999.  Both datasets were first 
processed through the standard WFPC2 pipleine at STScI.  
The IRAF/STSDAS task DRIZ$_{CR}$ was used to generate bad pixel masks for both static 
bad pixels and cosmic rays.  These masks were then applied and the images combined 
using the DRIZZLE routine with a scale factor of 0.5 and a pixfrac of 0.8.  To 
remove the diffraction spikes and other artifacts, the images were then deconvolved 
using 3 to 5 iterations of the the IRAF task LUCY utilizing synthetic TinyTim PSFs 
generated for each filter.  The PSFs were subsampled by a factor of two yielding a
scale of 0$\farcs$02275/pixel. They were then smoothed using a 3x3 to 4x4 pixel 
(5x5 pixel for the F1042M filter) boxcar smoothing algorithm. The short and long 
exposures were combined separately allowing for features to be seen in both the inner 
and outer regions of the nebulosity in spite of the high contrast between the two.
The combined short and long exposures for each filter are labeled {\it ``s''} and 
{\it ``l''}, respectively, see Table 1.  The two epochs of images, the {\it ``s''} and
{\it ``l''} combined exposures in each filter, were then rotated and aligned on the 
central star. Although, the centroid of the star shifts with wavelength (Kastner \& 
Weintraub, Smith et al 2001), by aligning the images from the same filters and 
with the same exposure times, we avoid the wavelength dependent shift.

\subsection{The Transverse Motions}

The angular expansion of  VY CMa's circumstellar nebula is not known. Maser
measurements (Bowers et al 1983; Richards et al 1998)  suggest an expansion of 35 km s$^{-1}$ near the star, while the long-slit
spectroscopy showed relative velocities and shifts of 50 -- 60 km s$^{-1}$ across 
the arcs. 
Assuming the latter, for the structures we want to measure, yields an expected
positional shift of 0$\farcs$051 in the 6.23 yrs between the two epochs. Although this
is small, the measurements are feasible with our pixel scale. 

To determine the transverse  motions and direction of motion  of discernible features between 
the two epochs, we initially examined each  pair of images  by eye in SAOImage DS9 to 
determine which features could be identified at both epochs.   The more prominent features 
were then 
labeled following the naming convention in Smith et al  and Humphreys et al, 
although it was necessary to
introduce some new identifications for weaker features.  Multiple positions were 
measured in some of the larger features such as the NW arc and they are numbered 
accordingly. The measured positions are identified in the accompanying images in 
Figures 1 -- 4.  Some features were more easily identified and measured in some filters than others. For example, the knots  close to the star are more easily viewed and 
measured in the blue exposures (F410M) while the dusty knots and filaments in  the far red 
F1042M filter have a much smoother, more amorphous  appearance making it more difficult to both
identify and measure features in it. 

Due to the high contrast within the nebula,  it was often necessary 
to adjust the brightness transform within DS9 to get a comparable contrast between 
the two epochs for each feature. Furthermore, it appeared that VY CMa had brightened
somewhat between 1999 and 2005, with some parts of the nebula especially near the star 
apparently 
brightening more than the outer nebulosity, thus supporting suggestions of variable illumination \citep{Wall78,Mon99}.  
The images were then blinked between the two epochs and the x and y pixel coordinates of the 
peak transmission of each recognizable feature were recorded.  This process was 
repeated for each feature in all of the combined short and long exposures in which
it could be identified.  Three independent measurements were made of each feature 
by working through 
the entire set of images before beginning the measurements a second and third time.  
The average of the net shifts in arcsec for the separate features is given in Table 2
for each filter and exposure time combination. The errors were determined from the 
standard deviation of the three measurements, and  are likely 
higher than any systematic errors arising from rotation, alignment and deconvolution.
Although our procedure adjusted for possible variable illumination, it may contribute
to the random errors. 
All of the measurements are included in Table 2, however for those features measured in 
three or more images, we could occasionally identify discrepant measurements or positions  
very likely due to the complexity of the ejecta and variable appearance of some of the
features in the different filters. These are identified in the footnotes to Table 2 and are  not 
included in the mean transverse velocity weighted by the standard deviation (Table 3).  
When a feature was measured in two
images where the positions in either epoch disagreed by more than 5 pixels, they are flagged
as separate features in the footnotes.

In Table 3 we give the projected radial distance and the position angle from the star in the 
plane of the sky for each measured feature  with its weighted mean transverse velocity 
(V$_{T}$), its  associated error in km s$^{-1}$ and its direction of motion ($\phi$), 
determined from the mean angular shift in the x and y coordinates between the two epochs with the error
of the mean. 
The vectors are shown as arrows on Figures 5 -- 7  corresponding
to the features  marked in the Figures 1 -- 4.  
In the next section we combine the transverse and Doppler velocities for a discussion
of the kinematics of the ejecta, the motions of specific arcs and knots, their ages
and orientations and the overall expansion of the nebula. 

\section{The Kinematics of the Ejecta}

Humphreys et al  obtained long-slit spectra at four positions across the nebula
shown in Figs.\ 1a and 1b in that paper.  Three
parallel slit positions  crossed the NW arc along the SE/NW direction and sampled 
clumps of knots and filaments closer
to the star.  The fourth, oriented roughly  NE--SW, crossed the two outer 
filamentary arcs 1 and 2.  Spectra were extracted from regularly-spaced apertures 
along the slits selected to cover specific features  as listed in Table 2 of the 2005 paper.
These data provide the  Doppler velocities discussed below for  positions
across the nebula and specific features in the WFPC2 images.

The most complete and precise Doppler velocities in these data are provided by  
the K I  $\lambda$7699 emission line,  which we attribute to   
resonance scattering at the observed locations. 
In principle the observed K I feature might represent
any of four processes:  (1) resonance scattering as just
mentioned;  (2) local thermal emission;  (3) potassium
recombination; or (4) emission or pseudo-emission produced near
the star and then reflected toward us by dust grains in the
observed regions.  For cases 1--3 the apparent Doppler shifts
have the usual meaning;  but case 4 involves the ``moving mirror''
effect which alters the relation between space motion and net
Doppler shift\footnote{Doppler velocities are available for a few strong absorption 
lines reflected by local dust but they involve the moving-mirror effect and are 
difficult to use.}.    Which is correct?  Some relevant quantities were
estimated in the first half of Appendix A in Humphreys et al. 
The scarcity of potassium makes cases 2 and 3 very unlikely
for such a prominent feature.  Possibility 4, reflection by dust,
is less likely because the measured line components at most of our measured positions
are sharper and less complex than those seen close to the star. Moreover, if
we interpret the measured Doppler shifts in the normal way as
velocities along the line of sight, then we find plausible and
self-consistent results  described below;  but this is not true
if we adopt the ``moving mirror'' formula instead.  We therefore
assume that the strong \ion{K}{1} $\lambda$7699 component measured
at each position is caused by local resonance scattering.
Strictly speaking we cannot {\it prove\/} this, but it is the simplest 
explanation, there is no apparent evidence against it,  and it leads to 
reasonable conclusions \footnote{
In the second half of Appendix A in Humphreys et al, we describe a 
 way to avoid a line-formation problem in the extremely strong K I emission
 line viewed {\it near the central star\/}.  That problem  does not
 arise in the material discussed in the present paper,
 thousands of AU from the star.  
 For our purposes here it is sufficient to view resonance
 scattering as simple reflection limited to the line's
 narrow velocity width in each local arc or condensation.}.

     If the 7699 {\AA} pseudo-emission feature observed in a condensation
     represents resonance scattering by K I, then one might ask
     whether its apparent wavelength is perturbed by K I 
     self-absorption along our line of sight through the surrounding
     material.  There are several arguments against that possibility.
     First, the spatial extent of each measured line component
     almost perfectly matches the corresponding feature in the
     WFPC2 images;  this can be seen in our 2D spectra, e.g.,
     Figs.\ 9 and 12 in Humphreys et al.   Second, the line component
     that we measured at most of the locations  is rather simple, not asymmetric
     or multiple as would likely occur with additional self-absorption.
     Third, as estimated in Appendix A of the 2005 paper, the line-center
     optical depths are not predicted to be large except within
     dense condensations;  potassium should be very thoroughly
     ionized.  In summary, there is no reason to think that significant self-absorption
     occurs.

Obviously, with only four slits the entire nebula was not covered, and consequently only about 
two--thirds of the features (40 out of 66) listed in the preceding tables have Doppler or 
radial velocities. Our quoted values for the  local line-of-sight velocity $V_z$
are all relative to the VY CMa's reference frame which we assume to be the K I emission Doppler velocity at the central star,
$\approx$ +41.0 km s$^{-1}$ .  For reasons indicated in
Humphreys et al, this may introduce an uncertainty of a few km s$^{-1}$
 but there is no evident way to determine the correction \footnote{VY CMa's
  expected systemic velocity is $\sim$ 37 km s$^{-1}$.}. 
The three components of the motions for these features are summarized in Table 4.
We combined our measured transverse velocity with the corresponding Doppler velocity (V$_{z}$) to
determine the total space motion (V$_{Tot}$) and the combined direction of motion, $\theta$.  
The slit and extraction aperture for the Doppler velocity from Humphreys et al  are
given in the comment column.  In those cases where multiple velocity components are present along the 
line of sight, an explanation is provided as a footnote to Table 4 or in the following discussion of the kinematics 
of the more prominent features.

\subsection{The ``Nebulous''  Northwest Arc}
In their analysis of the Doppler velocities for both the reflected absorption lines and the
K I emission, Humphreys et al reported a strong velocity gradient across the nebulous arc
to the northwest of the star. The broadened absorption lines were significantly redshifted
with a velocity difference of $\sim$ +50 km s$^{-1}$ relative to the star due to the moving
mirror effect. Although the direction of the flow was not known, the evidence from the line
widths, the P Cygni profiles, etc., suggested expansion velocites between 35 and 70  km s$^{-1}$
and that the motion of the material in the arc was mostly transverse at an angle of $\pm$ 
$\sim$ 20$\arcdeg$ with respect to the plane of the sky. 

We have measured 10 different positions along the NW arc. Combining the transverse velocities 
with the emission line velocities at the same positions we determine a mean total space motion
for the arc of 45.7 $\pm$ 4  km s$^{-1}$ at an angle $\theta$ of 22 $\pm$ 7$\arcdeg$  away from us along 
the line of sight. The results for the separate positions range from 28 to 69 km s$^{-1}$ and 
$\theta$ from 7.5 to 38$\arcdeg$. At its distance from the star and taking the mean projection
angle into account, the material in the arc was ejected about 500 yrs. ago. 

It is possible that radiative acceleration may have altered the $r/V_{Tot}$ ratios  used to estimate the 
time since the ejection. We can estimate this for the NW Arc whose mass is roughly 
known ($\sim 3 \times 10^{-3} M_{\odot}$ (Smith et al)). As seen from the star, the NW Arc 
covers less than 300 square degrees and therefore intercepts no more than 1\% of the 
total luminosity. The corresponding momentum flux would accelerate this mass at about
$10^{-4}$ cm s$^{-2}$ or $\approx$ 15 km s$^{-1}$ in 500 yrs. In that case,  
about one third of the observed outward speed may be due to
 post-ejection accelration which would reduce the age estimate from 500 to 400 
 yrs.  The real effect is most likely smaller because radiative acceleration would tend
 to disrupt the observed coherent structure of the arcs. Therefore, we expect the 
 effect  to be small on the estimated ages and have neglected it for the NW Arc and the 
 other features in the ejecta.

Humphreys et al  suggested that the section of the arc near the tip where
it appears to be bending back toward the star, may actually be the nearer side. Unfortunately,
there were no measurable knots or features in this section that were sufficiently resolved to 
confidently measure the motions. However, adopting our expansion velocity and the moving mirror velocity
relative to the star of 40 km s$^{-1}$ for the absorption lines at this location  
(Humphreys et al) we derive an angle $\sim$ -7.5$\arcdeg$ out of the plane, towards us, compared to 
+10$\arcdeg$, using the same method, for the same absorption lines along the major axis of 
the NW arc.
Thus this section of the arc does appear to be nearer.

\subsection{The Outer Filamentary Arcs 1 and 2} 

The K I emission lines across Arcs 1 and 2 show complex profiles and multiple peaks
due to  flows or streams of gas identified with these two very visible arcs.
( See the profiles in Figure 11 and the two-dimensional image of the slit in Figure 13 in 
Humphreys et al.) The extraction aperture across Arc 2 has  two very prominent emission
peaks. One  has a large positive velocity like that of the reflected absorption lines 
which we also attribute to reflection by the background material, while the blueward peak, 
which represents a separate flow of emitting gas along
the line of sight, is most likely produced by resonant scattering. We have therefore adopted this 
emission velocity for the radial component of the motions in Arc 2 which is -19 km s$^{-1}$ 
relative to the star. A strong blueshifted emission
feature which first appears  weakly in aperture 5 can be traced over 3$^{\arcsec}$ and across 
Arc 1. This is a kinematically separate flow of gas from that associated with Arc 2, and 
Humphreys et al identified this emission with resonant scattering by the gas in a flow 
associated with Arc 1. This gives a radial velocity relative to the star of -37 km s$^{-1}$. 

The transverse motions for the six  positions measured in Arc 2 show that they are moving in
different directions (Figure 6) consistent with an overall expansion of the arc like a bubble or
loop with a net motion to the south. Three of our positions correspond to the extraction aperture across Arc 2. Adopting the
corresponding radial velocity described above gives a mean space motion or expansion 
velocity of 64  $\pm$ 2.1 km s$^{-1}$  at an angle of -17$\arcdeg$, towards us
along the line of sight. At this angle and at a distance 3.4 arcsec from the star, the 
material in Arc 2 was ejected  460 yrs ago, about the same time as the NW arc.

Similarly for Arc 1, three of our measured positions with transverse motions have corresponding
radial velocities and give a mean expansion velocity of 68.2 $\pm$ 2.5 km s$^{-1}$
out of the plane of the sky towards us at an angle of -33$\arcdeg$.
The transverse motions measured at eleven different positions along the arc indicate an expansion of the loop with a net
motion of the material to the SSW on the sky. At its distance from the star and
moving at an angle of 33$\arcdeg$, Arc 1 was ejected about 800 years ago. 

Thus both Arcs 1 and 2 are most likely in the foreground, moving towards us, but at significantly 
different angles and in different directions.

\subsection{The W Arc}

Humphreys et al described this feature as a small irregularly shaped arc between the
star and the more prominent NW arc. Our measurements show that this feature is made up of
several small knots moving in an approximately northwest direction. Two separate slits cross the different knots or positions we measured
for the transverse motions ( Slit III ap 3 and Slit I ap 3), but like numerous places 
in the nebula, the K I emission line has more than one  velocity component in these apertures. 
After inspection of the line profiles and the 2-dimensional image of Slit III in Figure 9 in
Humphreys et al, we adopted the velocity peaks at 41.7 km s$^{-1}$ for III ap 3 and 45.6
km s$^{-1}$ for I ap 3. The latter is used only for knots D1 and D2 which based on their 
position may be distinct features separate from the W arc.  These adopted  velocities,
yield essentially zero radial motion relative to the star and a total space motion of $\approx$
44 km s$^{-1}$ to the northwest in the plane of the sky. The corresponding  time since the ejection
is $\approx$ 300 yrs.  

If we had chosen the slower radial velocity feature at $\sim$ 25 km s$^{-1}$, the knots in the
W arc would be moving towards us (-17$\arcdeg$) at $\sim$ 50 km s$^{-1}$ and ejected 
275 yrs ago. 
\subsection{The SW Knots}

Several small knots can be easily discerned in the WFPC2 images of VY CMa just to the southwest 
of the star.  Knots A,B, C and G in this grouping are 
 in the Slit III ap 2 extraction. We adopted the K I velocity component at 27.3 km s$^{-1}$,
yelding a radial velocity relative to the star of -14 km s$^{-1}$ for these knots. It is 
not clear if this clump of knots is gravitationally bound or moving together. However, if we 
treat them as a group, they are apparently moving out of the plane at -25$\arcdeg$ with a net
space motion of 36 $\pm$ 5.5 km s$^{-1}$. Their net transverse motion towards -86$\arcdeg$ is
essentially to the west, except for knot B. However, given the range in their
 vector motions, this feature could also be interpreted as a clump whose separate
knots are expanding away from each other. Assuming that they were ejected together at the same
time, it occurred about 250 yrs ago. If we had adopted the other velocity component,
(44.5 km s$^{-1}$) we would have concluded that the SW knots are moving essentially in the plane of the sky, 
but with little change in the expansion velocity (33 km s$^{-1}$) or the time since their ejection.

\subsection{The S knots}

This group of small knots directly to the south of the star is similar in appearance to
the SW group. Knots A, B, C, and Y are in Slit V ap 3 while knots D1 and D2 are in Slit III
ap 2. For the latter, we adopted the same velocity component used for the SW knots, while
at Slit V, one of the emission components is the same as that measured for the absorption
lines and is therefore associated with reflection from the surrounding material; see discussion
above for Arcs 1 and 2. 
The other much stronger emission peak at 37 km s$^{-1}$ gives a velocity relative to the star
of -17 km s$^{-1}$. The knots are thus in the foreground at an angle of -27$\arcdeg$. 
The S knots are moving to the southeast, position angle $\sim$ 156$\arcdeg$,  except
for knots D1 and D2 which given their position, may be part of a separate feature. The mean space motion 
or expansion velocity is 41.6 $\pm$ 5 km s$^{-1}$. This gives a time since the ejection of only 
157 yrs, the most recent age among the variety of embedded structures that we have discussed. 
With an uncertainty of $\pm$ 25 yrs for the mass loss episode, the S knots, and the 
subsequent formation of dust, may correspond to VY CMa's fading from $\sim$ 6.5 mag to 8 mag 
from $\sim$ 1870 to 1880.

\subsection{The S Arc}

The S Arc stretches approximately east to west about 2$\farcs$5 to 3$\farcs$0 from the
central star. Slit V crosses knot B in the S arc between apertures V ap4 and V ap5. with 
three and two emission peaks in each aperture, respectively. We associate the most 
redshifted velocity with background material as discussed for Arc 2. This leaves K I 
emission features at 25 km s$^{-1}$ and 52 km s$^{-1}$ for possible kinematic identification 
with the material in the S Arc. Their respective velocities relative to the star are
-16 km s$^{-1}$ and +11 km s$^{-1}$ corresponding to -22$\arcdeg$ and +16$\arcdeg$ for
the S Arsc's orientation with respect to the plane of the sky. We favor the foreground
orientation based on the appearance of the S Arc, but either is possible. The total 
space velocity is $\sim$ 41 -- 42  km s$^{-1}$, and the time since the ejection 
is 480 yrs in both cases.

\subsection{The SE Loop}

We measured transverse motions at three positions on the  small SE Loop or arc on the east side 
of VY CMa's asymmetric ejecta.  Slit III
crosses knot B on the SE loop, but Humphreys et al did not include an extraction at that position
which is near the  end of the slit. We therefore went back to the original 
spectrum and measured a Doppler velocity at that position of + 17.5 km s$^{-1}$, 
correspondig to a velocity relative to the star of -23.5 km s$^{-1}$.
The SE Loop is therefore in the foreground at an angle of -21$\arcdeg$  moving
to the southeast. The space motion for knot B is 65.1 km s$^{-1}$, yielding an age
of $\sim$ 320 yrs. The SE Loop is the only feature on the east of the 
ejecta for with  measured transverse motions. There were no easily measured knots in 
``arc 3'' identified by Smith et al also on the east side of the nebula.

\subsection{The SW Clump}

What we have called the SW Clump is one of the more perplexing features in the ejecta.
It is close to the star, but only seen in the far red F1042M filter and is located 
between the SW and S groups of knots. It is obviously very red and dusty.
Two of the long slit  extraction apertures overlap the SW Clump, but since this feature 
is totally obscured at the shorter
wavelengths, it is uncertain if the measured radial velocities at this position are applicable; 
although, the K I emission line is in the far-red at 7700$\AA$. Since the SW Clump is highly obscured it seems
reasonable to assume that it is not a foreground feature, so we have adopted the more
redshifted velocities in these two apertures of 42 - 44 km s$^{-1}$ which gives a 
negligible radial velocity with respect to the star. If this is the case, the
SW clump is moving slightly away from us ($\sim$ +8$\arcdeg$) quite slowly at only $\approx$ 18 km s$^{-1}$ to the SSW, and  was ejected about 500 yrs ago. 

\subsection{General Expansion of the Nebula}

Over much of the extended nebula, the K I emission line has a velocity component
 at or near its heliocentric velocity measured at the star 
(41.0 km s$^{-1}$)  and the
expected systemic velocity of $\sim$ 37 km s$^{-1}$ (Bowers et al 1983).
Indeed, in the outer parts of the nebula, the K I emission line is often double, 
showing both of the above velocities (Tables 4 and 6 in Humphreys et al).
The K I lines in the outer ejecta are also very narrow (10 --15 km s$^{-1}$ ), indicative of very
little Doppler broadening. Furthermore, the H$\alpha$ line, which is quite weak on
and near the star, becomes relatively strong in the outer parts of the nebula which
Humphreys et al attributed to nebular emission from the nearby H II region (Sharpless
310). Its  velocity agrees with the systemic velocity and with the same
 velocity feature in the K I emission line. Thus we also attribute the latter to 
reflection by dust in the surrounding medium, not to VY CMa's ejecta. 
This still leaves the K I emission near 40 km s$^{-1}$ which if due to reflection by
the ejecta, implies virtually no radial expansion relative to the star. 
Thus any measurement of transverse motion in the diffuse nebulosity (separate from
the arcs and knots) is important to determine any overall expansion of the nebula.

Unfortunately, this proved difficult due to the lack of measurable features in the 
more diffuse outer ejecta. We did measure the transverse motions for several ``spikes'' or
extensions at the western edge of the visible nebulosity; some of which had quite high
tranverse motions (Table 3). Two of these also had measured Doppler velocities at 35 - 42 
km s$^{-1}$; thus their radial motion relative to the star is $\sim$ 0 km s$^{-1}$ and
their total space motion is nearly all transverse at 30 - 40 km s$^{-1}$. At 6$\arcsec$ to 7$\arcsec$
from the star this material would have been  ejected about 1300 to 1700 years ago. 

Humphreys et al also noted  broad wings on the K I emission lines in the outermost
ejecta along all four slit positions. The wings have Doppler velocities which average
-136 km s$^{-1}$ and +208 km s$^{-1}$ at positions between 7$\arcsec$ and 9$\arcsec$ from the star.
They  very likely represent faster moving, more diffuse gas and dust perhaps from an earlier more 
uniform ejection or wind.

\section{Discussion -- Geometry of the Ejecta}

The results for the spatially recognizable features discussed in the preceding section
are summarized in Table 5. Except for the NW arc and perhaps the SW Clump, all of the 
measured features appear to be moving either close to the  plane of the sky or 
toward us. 
This is not surprising, since within about 3$\arcsec$ of the star, most of the diffuse  nebulosity  appears to be optically thick. Evidently, we do not see through 
the nebula and do not  see many features, presumably on the other side,  
moving away from us. Figures 8 -- 10 show 3D representations of the positions of the 
knots for which we have total space motions. The figures show their positions relative
to the plane of the sky assuming uniform radial expansion viewed from three 
different perspectives \footnote{This 3D visualization can be viewed
as a movie at www.astro.umn.edu/$\sim$ahelton/research/VYCMa.}. 
Our results for the vector motions of the major arcs and 
the clumps of knots clearly show that these structures were not only ejected at 
different times, but are also moving in different directions and at different angles
relative to the plane of the sky and to the star. This is definitely suggestive of  
random  locations for the sites of the 
ejection episodes on the star and the directions of the outflows.

The extensive maser (OH. H$_{2}$O, SiO), and CO observations however, have been 
interpreted as evidence for  an axis of symmetry with  possible bipolar outflows and 
a disk-like distribution for the circumstellar material, although, the models for the 
geometry of the system based on the different 
maser distributions and CO maps, are not always consistent with each other. 
Most of the maser spots and the intensity maxima  are also quite close to the star, 
typically within 0$\farcs$5, consequently there is
little  correlation with the optical knots and arcs discussed in this paper. Only our NW knot is this
close to the star, but its position does not appear to correspond to any of the maser 
emission spots. 

The OH maser intensity maxima are within 0$\farcs$5 and   
 are distributed along a NE-SW axis at a position angle of
$\sim$ 50$\arcdeg$ (Bowers et al 1983) which could be tilted 15 to 30$\arcdeg$ to our
line of sight. If there is a NE-SW polar/rotation axis, then arcs 1 and 2 could both
be part of an associated bipolar outflow depending on the opening angle of the cone, but 
from different locations on the star corresponding to the separations of their vector 
motions of $\sim$ 15\arcdeg. In this case, as Humphreys et
al suggested, the NW arc would then be near the corresponding equatorial plane. 
With this assumed geometry, the three prominent arcs could correspond to ejection episodes near 
the polar and equatorial axes, but the S and SW clumps and other smaller 
arcs would be ejected from more random directions. 

The H$_{2}$O maser spots appear to be  oriented much more east-west (Richards, Yates \& Cohen 1998). 
The highly polarized   SiO maser spots (Shinaga et al 2004) have a 0$\farcs$2 north-south distribution 
on the 
sky, although  their polarization vectors  have a mean position angle  of 72$\arcdeg$ in agreement with 
the orientation of the bipolar axis of the SiO emission. Similarly, diffraction -- limited speckle 
interferometry in the near -- infrared also shows a primarily north-south orientation for the 
dust shell extended 0$\farcs$2 with a position angle $\sim$ 153 to 176$\arcdeg$ 
(Wittkowski, Langer \& Weigelt 1998), while near--IR aperture--masked interferometry
 (Monnier et al 1999) showed a  southward extension of the dust emission within 0$\farcs$1
but  no clear disklike or bipolar geometry in the images.  
K-band interferometric measurements however, 
indicate a bipolar distribution within 0$\farcs$1 of the star but with the dusty
disk oriented east-west (Monnier et al 2004). 
Recent interferometric millimeter observations of CO and SO (Muller et al 2006) have been modelled
as a bipolar outflow in the east-west direction, but with a very wide opening angle ($\sim$ 120\arcdeg)
and an expanding shell elongated north-south. 

With this lack of strong evidence for a well-defined   or preferred axis of symmetry or bipolar axis
in VY CMa, we conclude that the kinematics  and corresponding morphology
of the  numerous arcs and knots are more consistent with {\it a 
history of localized mass ejections from active regions on the star not strongly aligned with a presumed axis or equator}.  

VY CMa's asymmetric circumstellar nebula, extended to the west and south of the star, 
is very apparent in the  optical images. The more prominent embedded
arcs and knots are also found to the west and south of the star, although the 
smaller SE loop and arc 3 (Smith et al 2001) are to the east of the star. 
The nebula appears much more symmetric in the infrared with more symmetric contours at 
2$\mu$m and in the thermal infrared (see Figures 4 and 5 in Smith et al). 
 Smith et al also showed that the reddening and apparent extinction was much
 higher to the east and north of the star. They suggested that a 
 combination of higher extinction plus possible  back-scattering, assuming a 
 NE-SW axis of symmetry, could account for the lack of
 visible nebulosity to the east and northeast. This interpretation may be supported 
 by the contours at 9.8$\mu$m which appear to be compressed or foreshortened to the 
 northeast. We also want to point out, however, that VY CMa is on the western 
 edge of the large dark cloud Lynds 1667 and at the same distance. Thus its 
 asymmetric appearance could be due entirely to obscuration.

\section{Conclusions -- Convective Activity, Magnetic Fields and Mass Loss}

We first suggested in Smith et al that the complex ejecta of arcs and knots
revealed in the first epoch HST/WFPC2 images resulted from ejection episodes,
possibly involving large scale convection and magnetic fields. We also demonstrated
that the arcs are too massive to have to have been ejected by radiation pressure alone 
and that the initial ejection was caused by some other process. The measured Doppler 
velocities in Humphreys et al showed that the arcs were kinematically separate
from the surrounding diffuse material and represented separate gas flows expanding
relative to the star. With the addition of the transverse motions reported here,
ages, velocities,  and directions of the outflows confirming their origin from 
eruptions at different times and from spatially separate  regions on the star not 
by a more uniform long-term mass loss. 

Smith et al estimated the mass of the NW arc to be $\sim$ 3 $\times$ 10$^{-3}$M$_{\odot}$ from its surface brightness
in a two arcsec$^{2}$ section, assuming an optical depth of unity. This is likely an
underestimate because the NW arc is may be optically thick.  Similarly, Arcs 1 and 2 
are also the visible loops or bubbles produced by large ouflows of gas extending over 
several arc seconds on the sky (Humphreys et al), and are probably  as massive as the NW
arc.  Assuming 3 $\times$ 10$^{-3}$M$_{\odot}$ in  each of these three arcs plus the 
S and SW clumps of knots, there may be  more than $\approx$ 1.5 $\times$ 10$^{-2}$M$_{\odot}$  in these featues. This is $\approx$ 10\% of the total mass of 0.2 --0.4 M$_{\odot}$ 
(Smith et al) in the nebula, and this does not take 
into account the numerous small filaments and knots visible throughout the nebula or 
similar ouflows not visible through the  nebula. With a dynamical time scale of 3 yrs,
the short--term mass loss rate associated with the NW arc and
similar features is
$\sim$ 10$^{-3}$M$_{\odot}$ yr$^{-1}$, several times the  average mass loss rate.

The expected duration for a convective event or nonradial pulsation would likely be on
the same order as VY CMa's dynamical timescale of 3 yrs or perhaps slightly longer. 
If each of the more prominent features, the arcs and clumps of knots represents
a temporary mass loss of the order of 3 $\times$ 10$^{-3}$M$_{\odot}$, then the
total kinetic energy in the NW arc, for example, expanding at 46 km s$^{-1}$ is 
6 $\times$ 10$^{43}$ ergs.  This is modest compared to the $\sim$ 2 $\times$ 10$^{47}$ ergs VY CMa would radiate in 3 yrs, and is also comparable to the thermal
energy in the ejected mass. 

In our two previous papers on VY CMa we have suggested that the expanding arcs, loops,
and clumps of knots are the result of localized activity on the star related to
convection and magnetic fields. Nonradial pulsational instability may be an alternative
ejection mechanism, but the distinction may be vague for a red supergiant where the
convective cells are expected to be comparable to the stellar radius in size 
(Schwarzschild 1975); although, nonradial pulsations would not be expected to produce
the narrow arcs and loops observed in VY CMa. Starspots and large ``asymmetries'' have 
now been observed  
on several stars including red giants, AGB stars and supergiants. The best example
among the red supergiants is probably $\alpha$ Ori (Gilliland \& Dupree 1996), but 
stellar hotspots have also been observed on $\alpha$ Sco and $\alpha$ Her with 
properties consistent with a convective origin (Tuthill, Haniff \& Baldwin 1997).
Monnier et al (2004) have reported on high resolution imaging of evolved
M stars including NML Cyg, VX Sgr and VY CMa revealing large-scale inhomogeneities and deviations from uniform brightness
which they  attribute to magnetic fields and/or rotation.  

Recently, Vlemmings et al (2002, 2004) have estimated the magnetic field strength from the circular
polarization of H$_{2}$O masers in the ejecta of AGB stars and several
evolved supergiants including the strong OH/IR sources VY CMa, VX Sgr, NML Cyg, and S Per. They report magnetic fields in VY CMa of $\sim$ 200mG
at distances of 220 AU. Their analysis supports the Zeeman interpretation of the circular
polarization of the SiO masers only a few AU from the surface of the stars (Barvainis et al 1987; Kemball \& Diamond 1997). Together
with Zeeman splitting of the OH  emission far out in the wind at a few thousand AU (Szymczak \& Cohen 1997;  Masheder et al 1999), these 
measurements confirm the presence of a magnetic field throughout the ejecta of VY CMa. 
Each of these results  imply magnetic fields of the order of  10$^{4}$G  at the star's surface assuming the r$^{-2}$
dependence of a solar--type magnetic field that would be associated with large star spots,
convective activity and the mass ejections. 

VY CMa is a member  of a relatively small group of evolved, highly unstable, massive stars called cool hypergiants, 
that are just below the   empirical upper luminosity boundary in the HR Diagram. Among 
this high luminosity group a few stars stand out, the OH/IR supergiants mentioned above 
plus IRC+10420, with exceptionally high mass loss rates, 
and resolved circumstellar ejecta (Humphreys et al 1997, 2002, Schuster, Humphreys \& 
Marengo 2006). There is no evidence for a close companion in any of these stars which 
could be responsible for their mass loss and ejecta (see Smith et al for VY CMa). These stars  may represent a short-lived stage with episodes of high mass loss.  We are thus observing increasing evidence among the evolved massive stars
($\eta$ Car, LBVs, and the cool hypergiants) for episodic mass loss. In the cool 
hypergiants (IRC+10420, VY CMa, VX Sgr, S Per, NML Cyg), the high mass loss episodes 
may be driven  by large -scale convection and magnetic fields. 
In Paper II we present the polarimetry and the three dimensional spatial structure 
of VY CMa's circumstellar ejecta. 

\acknowledgments
We thank Kris Davidson for many helpful discussions and George Wallerstein for reading 
and commenting on a draft  of the manuscript.  It is always a pleasure to thank 
George Herbig  for his support and continued interest in VY CMa.
This work was supported by NASA through grant number GO10262 from the Space Telescope
Science Institute.  

{\it Facilities:}  \facility{HST (WFPC2)}   \facility{HST (ACS)}

\appendix

\section{The Luminosity of VY CMa and Its Position on the HR Diagram}

The luminosity of  VY CMa ( $L \sim  4 - 5 \times 10^{5} L_{\odot}$
) is well-determined from its spectral energy distribution 
and distance, and places it near the empirical upper luminosity limit 
in the HR Diagram for cool hypergiants. 
In a recent paper however,  Massey, Levesque \& Plez (2006) suggest that  
VY CMa is nearly a factor of 10 less luminous than has previously been determined  
by several authors. They assert that its high luminosity and 
other ``extreme'' properties such as its inferred large size were based 
on an adopted effective temperature that was too low (e.g. $\sim$ 2800$\arcdeg$, Le Sidaner 
 \&  LeBetre (1996)).

 Massey et al fit  recent optical spectrophotometry of VY CMa 
 with  MARCS model atmospheres and derived a much warmer effective temperature.  
 Combining  its apparent visual magnitude, an adopted 
 interstellar extinction, and a temperature dependent bolometric correction,  
 they derive a luminosity $L  \sim 6  \times 10^{4} L_{\odot}$ 
 instead of the usually quoted  $\sim 4 -- 5 \times 10^{5} L_{\odot}$. 
 However, this classical  approach ignores one of VY CMa's  distinguishing 
 characteristics, its spectral energy distribution and large excess 
 radiation in the  infrared.

 A recently published example of its energy distribution can be seen in Figure 7 in Smith et al.
Most of the star's radiation is reprocessed by the dust in its extensive 
circumstellar ejecta. Its  energy distribution rises rapidly in the infrared and has a broad 
 maximum between 5 and 10 $\mu$m. Combining the photometry in Tables 3 and 4 in Smith et al 
for the entire nebula  with the IRAS data from 25 to 100 $\mu$m, and integrating the apparent
energy distribution, yields a luminosity of $L  = 4.3 \times 10^{5} L_{\odot}$ 
  at VY CMa's distance of 1.5 kpc (Herbig 1972, Lada \&  Reid 1978, Marvel 1997, 
   the same distance used by Massey et al.) If corrected for {\it interstellar} extinction at visual and red wavelengths, the luminosity would  increase by only a few percent,
because most of the most of the flux is escaping at $\sim$ 10$\mu$m. 
 Furthermore, an A$_{v}$ of 3.2 mag (Massey et al)  implies that at least 2 mag of 
  more of circumstellar extinction is required in the visual to equal  the flux emitted at 10$\mu$m. The wavelength dependence of the CS extinction correction, however, is not known.

The  standard ``textbook'' approach, relying only on visual photometry and
spectroscopy and  an  assumed temperature, is not valid for stars 
with sufficient circumstellar dust to reradiate their visual and red flux 
in the thermal infrared. In some cases, the radiating dust also dominates  the  
observed flux  between   1 $\mu$m and 5 $\mu$m and contributes significant circumstellar 
extinction  at visual, red and near-infrared wavelengths.
Other well-studied examples in 
our galaxy are VX Sgr, S Per, and NML Cyg. Like VY CMa, all three are strong
maser sources and NML Cyg is optically obscured. See Schuster, Humphreys \&  Marengo (2006) 
for recent images of these stars. 

In summary, the luminosity proposed for VY CMa by Massey et al is far less than what is
actually observed, and there is little doubt that it is near the empirical upper luminosity 
limit in the HR Diagram for the cool hypergiants (Humphreys \&  Davidson 1979, 1994).

Further consideration of VY CMa's  exact  position on the HR Diagram  depends on the assumed 
surface temperature. Previously published spectral types for VY CMa in the past 30 years
or so have been mostly in the M4-M5 range; although, Massey et al suggest that 
VY CMa's apparent spectral type is more likely $\sim$ M2.5 based on the 
MARCS model atmosphere fit to their spectrum. 
Interestingly, though, the blue TiO bands  in their published spectrum 
(Figure 2 in Massey et al) are more like  
their M4-type  reference spectrum than the M2-type spectrum they show. This author's 
numerous spectra of VY CMa obtained over many years have all been in the M4-M5 range. 
Therefore, adopting the M4--M5  spectral type with the temperature scale proposed by 
Levesque, et al (2005) 
gives T$_{eff}$ $\sim$ 3450--3535$\arcdeg$, while an older scale (used in Humphreys \&  McElroy
1984 from Flower 1977) yields T$_{eff}$ $\sim$ 3200$\arcdeg$ for an M4-M5 star.

However, one should be cautious in the case of VY CMa; 
{\it we are not observing either its 
photosphere or its surface directly.} It has been known for some time that VY CMa's absorpti
on
spectrum is significantly redshifted with respect to its systemic velocity
(Humphreys 1975, Wallerstein 1977) due to scattering by dust (Herbig 1970,
Kwok 1976, Van Blerkom \&  Van Blerkom 1978).
Indeed, most of VY CMa's visual-red radiation originates by reflection and scattering
by the dust grains at 100 AU from the star, the dust formation radius. Only a few percent 
of the  radiation actually escapes through the dust shell, which is very likely
inhomogeneous,   
implying optical depths of 4 to 5 at $\sim$ 7000$\AA$ in its wind (Humphreys et al).
If the wind is opaque, then R$_{ph}$ where the photons arise, could be larger than the
true stellar radius, and the underlying star possibly somewhat
warmer.

Massey et al also suggest that with previous temperature estimates, VY CMa would
violate the Hayashi limit.   
But the cause of the apparent conflict with the Hayashi limit is the assumed temperature
not the luminosity.
Whether or not it violates the Hayashi limit depends on whether 
the adopted temperature, inferred from the strength of the TiO bands or an 
atmospheric model, is indicative of the star's ill-defined surface or its
wind. With the above temperatures,  VY CMa is on the edge or just inside the 
Hayashi limit as plotted in Figure 1 in Massey et al., but the standard Hayashi 
limit applies to hydrostatic atmospheres.
Non-spherical outflows and a resulting dense wind as in VY CMa may affect the
Hayashi limit's location on the HR Diagram. 

VY CMa's high luminosity and apparent low temperature suggest that it is one of the largest 
stars known. Monnier et al (2004) derived a radius of 3000 R$_{\odot}$ from 2 $\mu$m
interferometry.  Given the above arguments,   3000 R$_{\odot}$ is probably not the actual 
size of the imbedded star. 
 Adopting this radius with VY CMa's luminosity gives an ``effective'' temperature of 
 $\sim$ 2700$\arcdeg$  which is rather low. Alternatively,   
 with the  apparent temperatures given above,  the radius is 1800 to 2100 R$_{\odot}$.  
 In either case,  {\it VY CMa is obviously very luminous, cool and big.}

\begin{figure}
\figurenum{1}
\epsscale{1.0}
\plotone{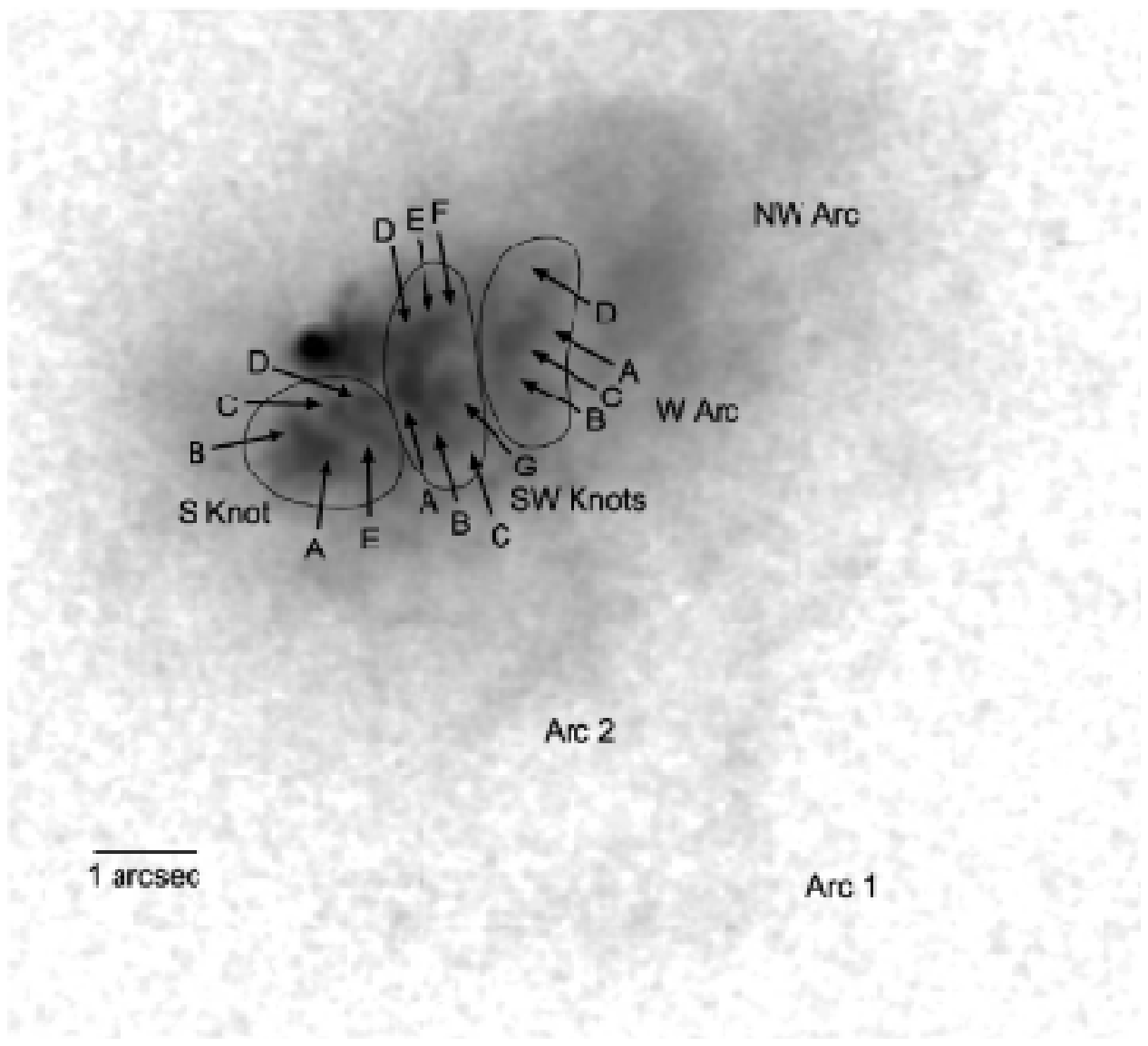}
\caption {The F410M image showing the  features in the inner ejecta including the S and SW knots.}  

\end{figure}

\begin{figure}
\figurenum{2}
\epsscale{1.0}
\plotone{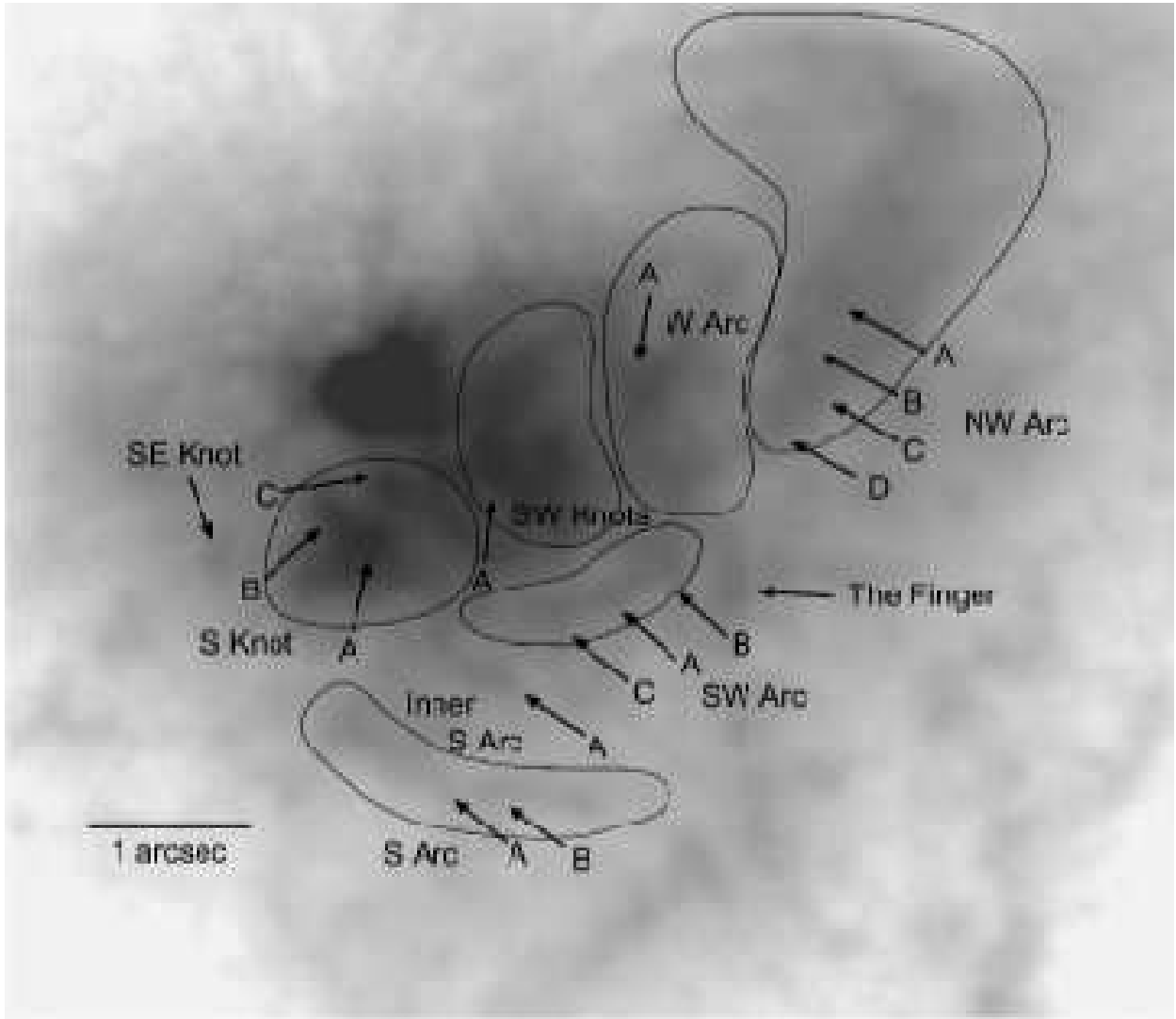}
\caption {The F547M image showing measured features in the ejecta including positions in the NW arc.}
\end{figure}

\begin{figure}
\figurenum{3}
\epsscale{1.0} 
\plotone{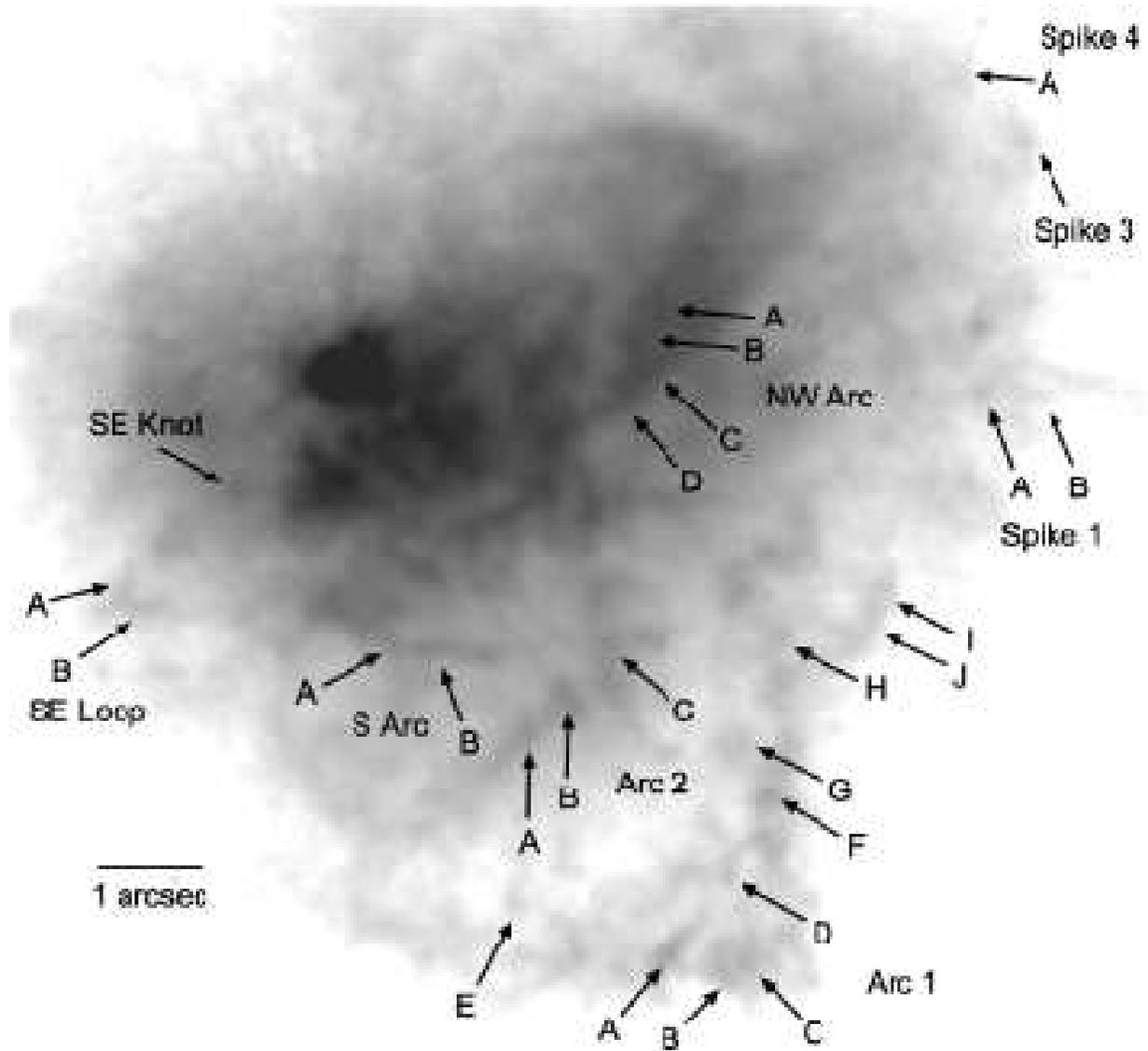}
\caption {The F547M image with positions in the outer ejecta dentified including Arcs 1 and 2.} 
\end{figure}

\begin{figure}
\figurenum{4}
\epsscale{1.0}
\plotone{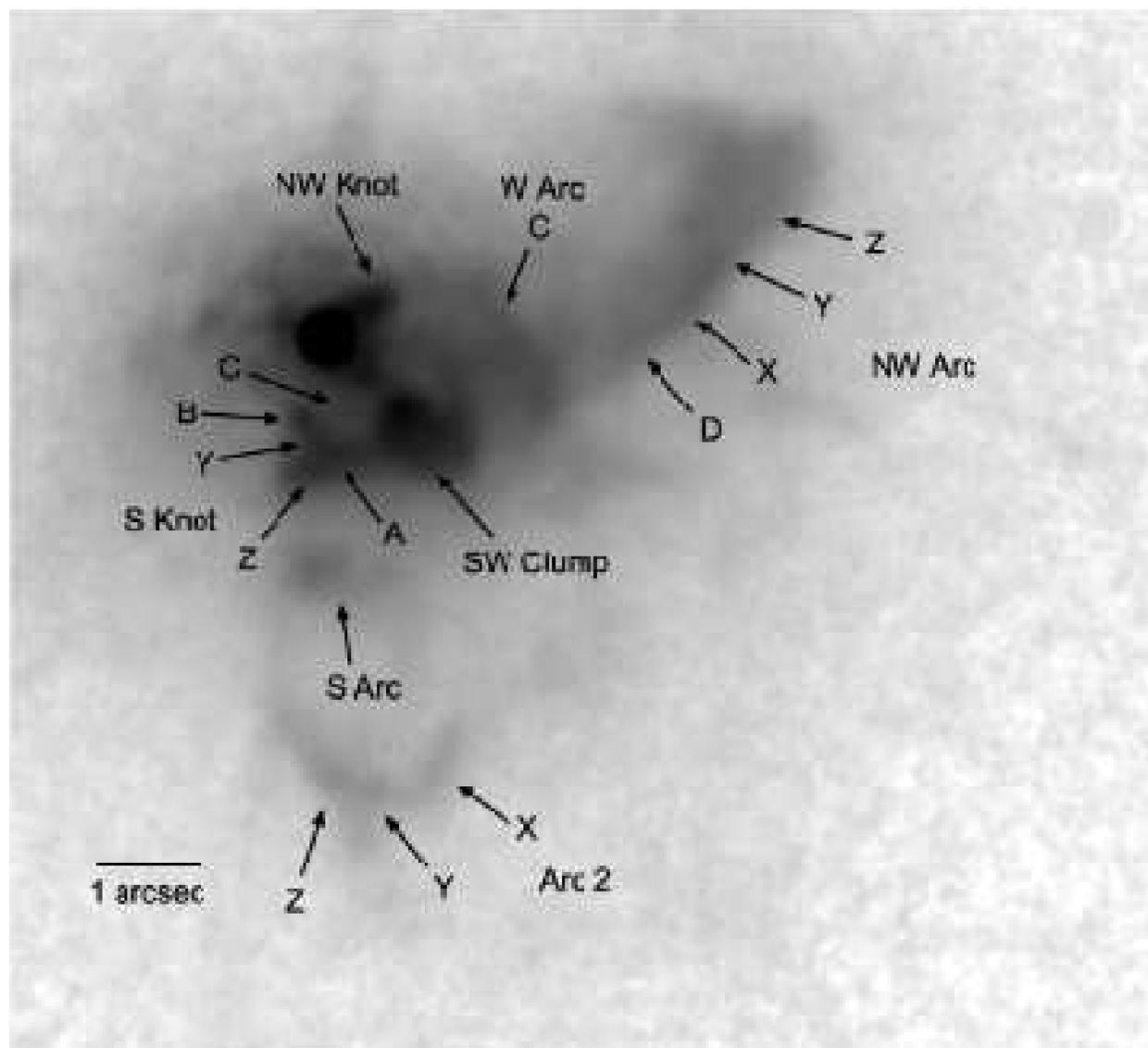}
\caption {The F1042M image showing the features measured on  this far-red image}
\end{figure}

\begin{figure}
\figurenum{5}
\epsscale{1.0}
\plotone{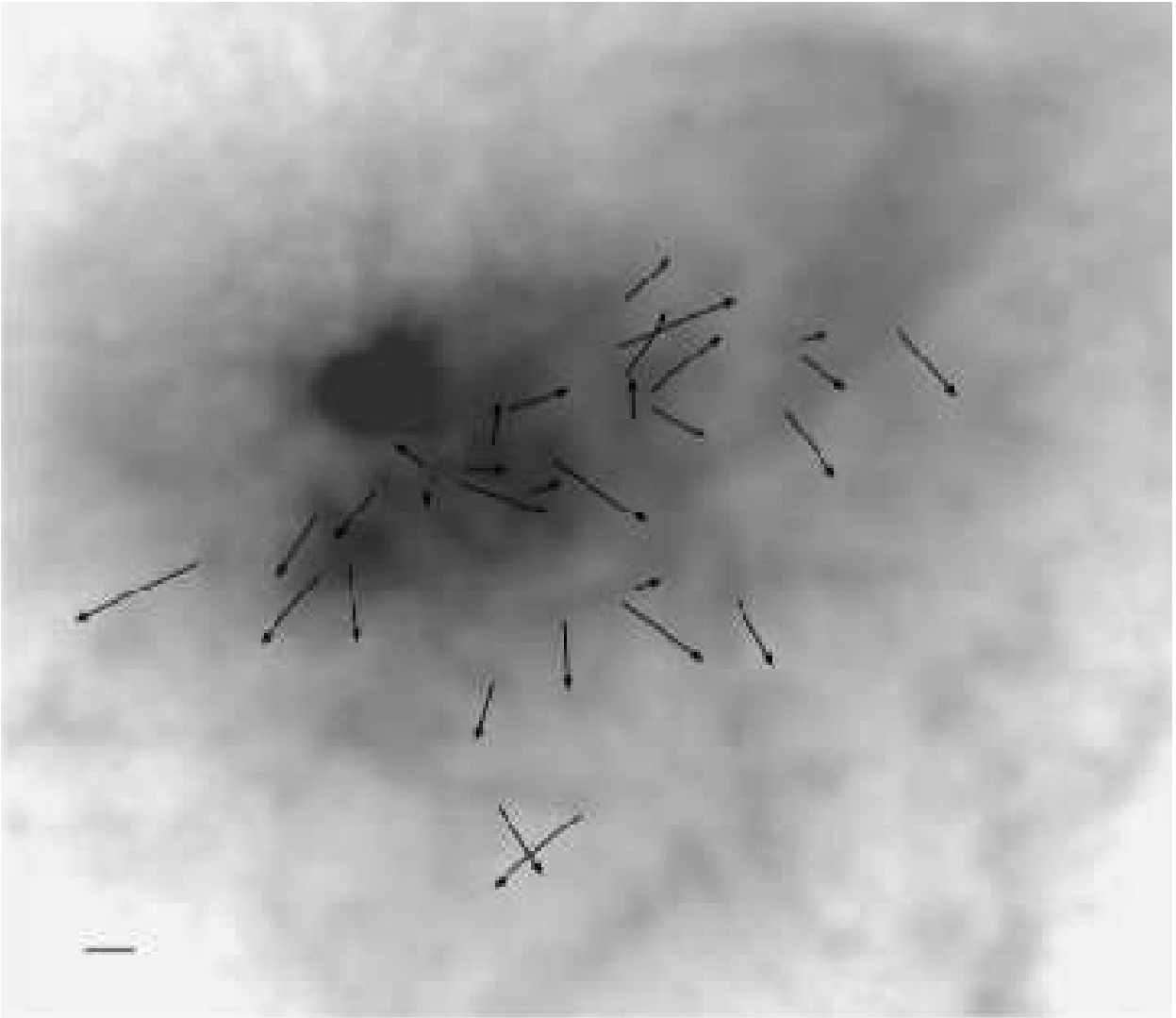}
\caption{The F547M image with the velocity vectors ($\phi$)for the transverse motion from Table 3. The length of the arrow is proportional to the transverse velocity; the scale bar in the lower left corner is 30 km s$^{-1}$. The features identified in Figures 1 and 2  are shown here.} 
\end{figure}

\begin{figure}
\figurenum{6}
\epsscale{1.0}
\plotone{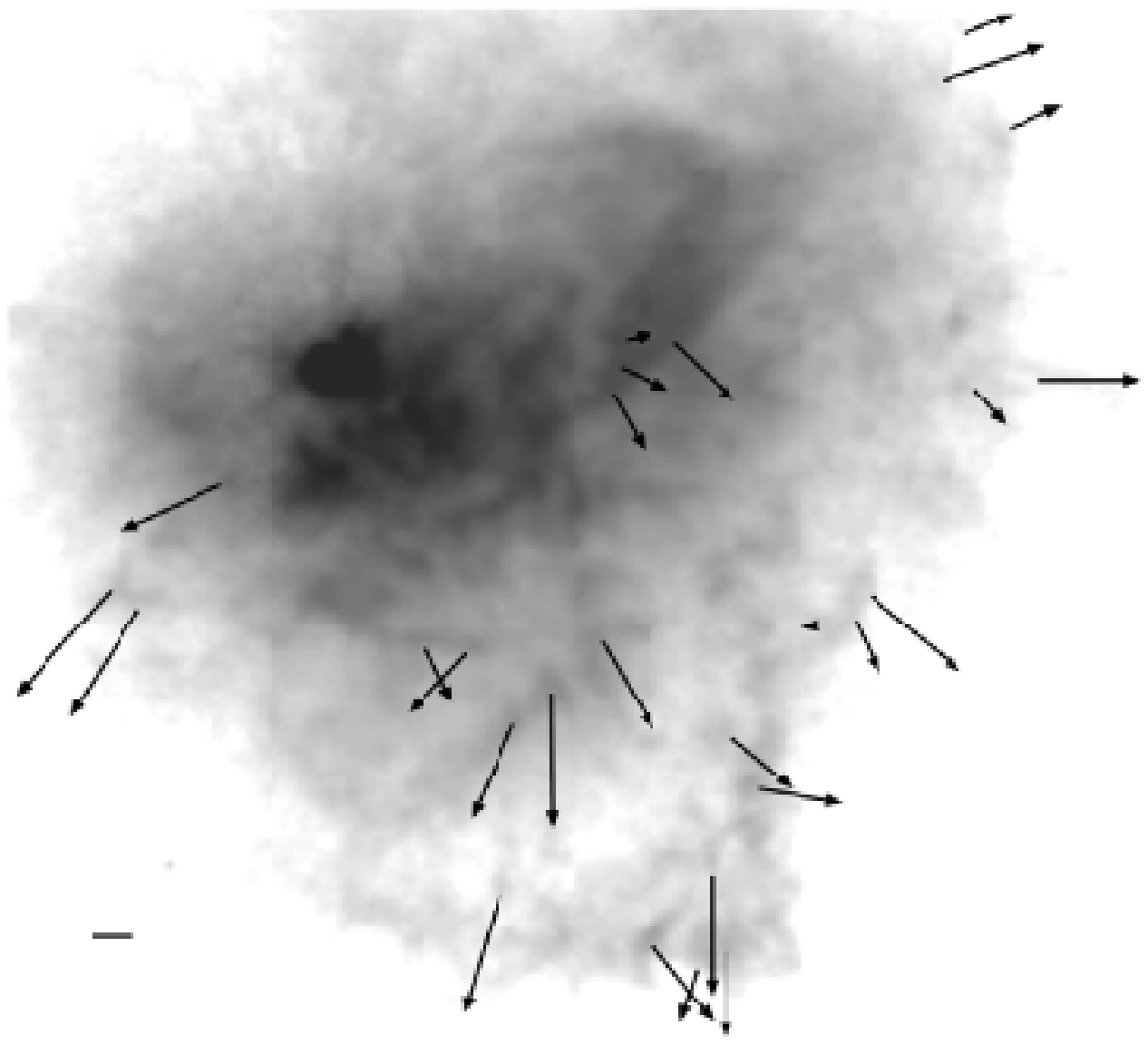}
\caption{Same as Figure 5 for the features identified in Figure 3.}
\end{figure}

\begin{figure}
\figurenum{7}
\epsscale{1.0}
\plotone{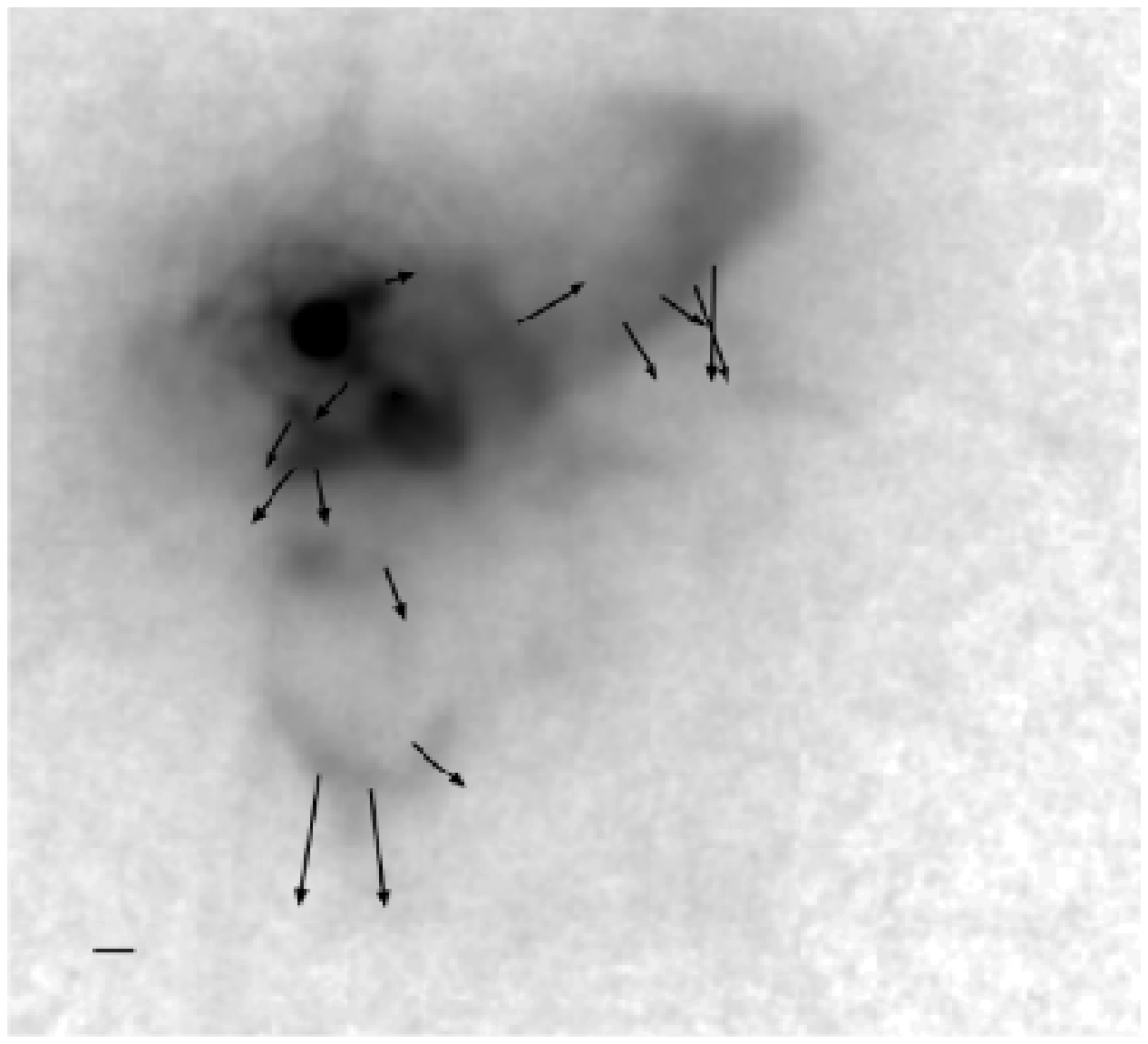}
\caption{Same as Figure 5 for the features identified in Figure 4.}
\end{figure}

\begin{figure}
\figurenum{8}
\epsscale{1.0}
\plotone{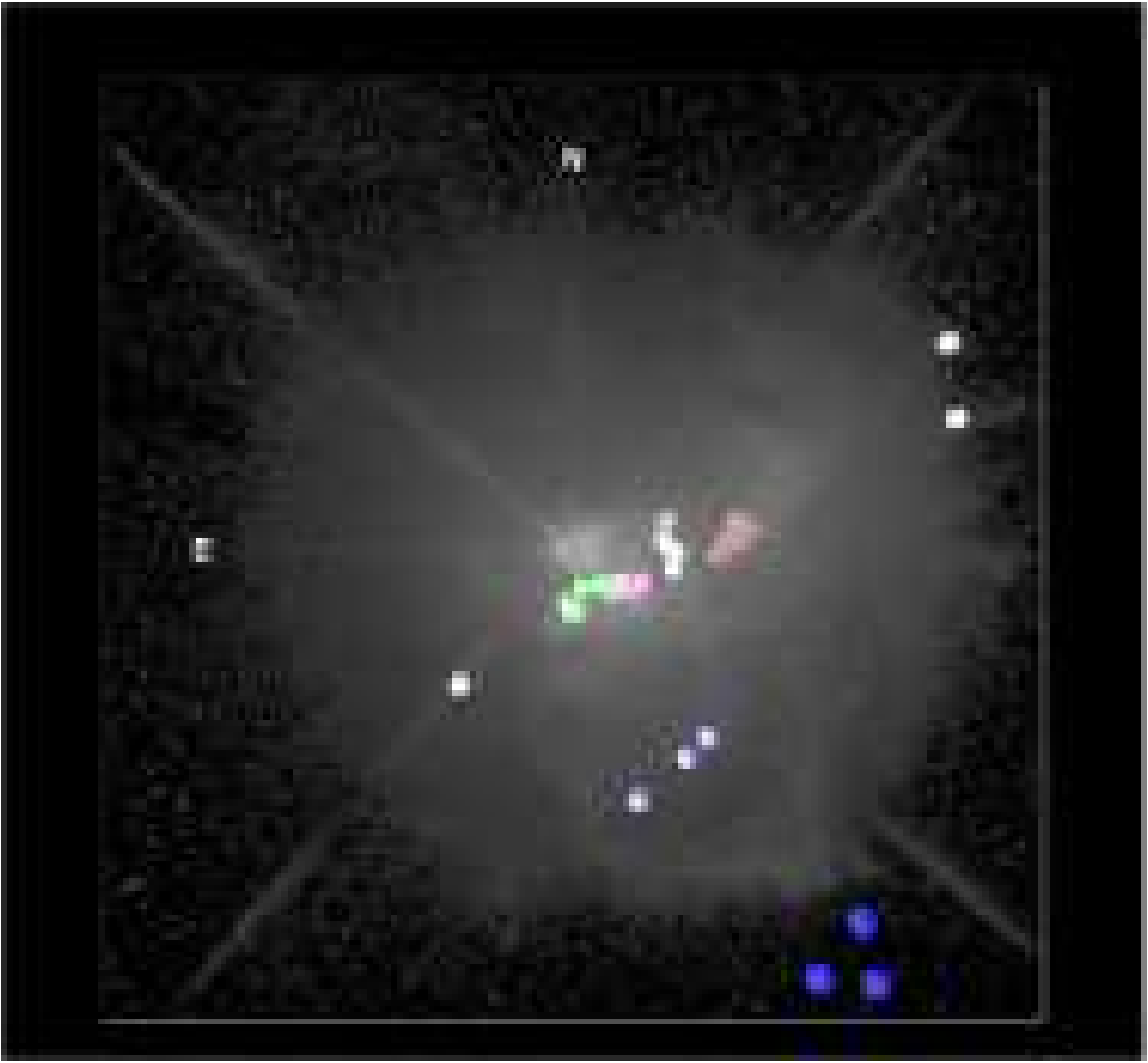}
\caption{The distribution of the measured knots with total space motions and 
therefore positions in all three dimensions  on an image of VY CMa. 
This is not an orthographic projection. In this projection the viewer is close 
to the star, hence, nearer objects (Arc 1) appear larger and at a wider angle 
relative to the star. The measured positions are color-coded: dark and light
blue for Arcs 1 and 2, respectively, orange for the NW Arc, green for the S lnots, pink for the SW knots and white for the W Arc, SE Lop, and `spikes' 3 and 4.}  
\end{figure}

\begin{figure}
\figurenum{9}
\epsscale{1.0}
\plotone{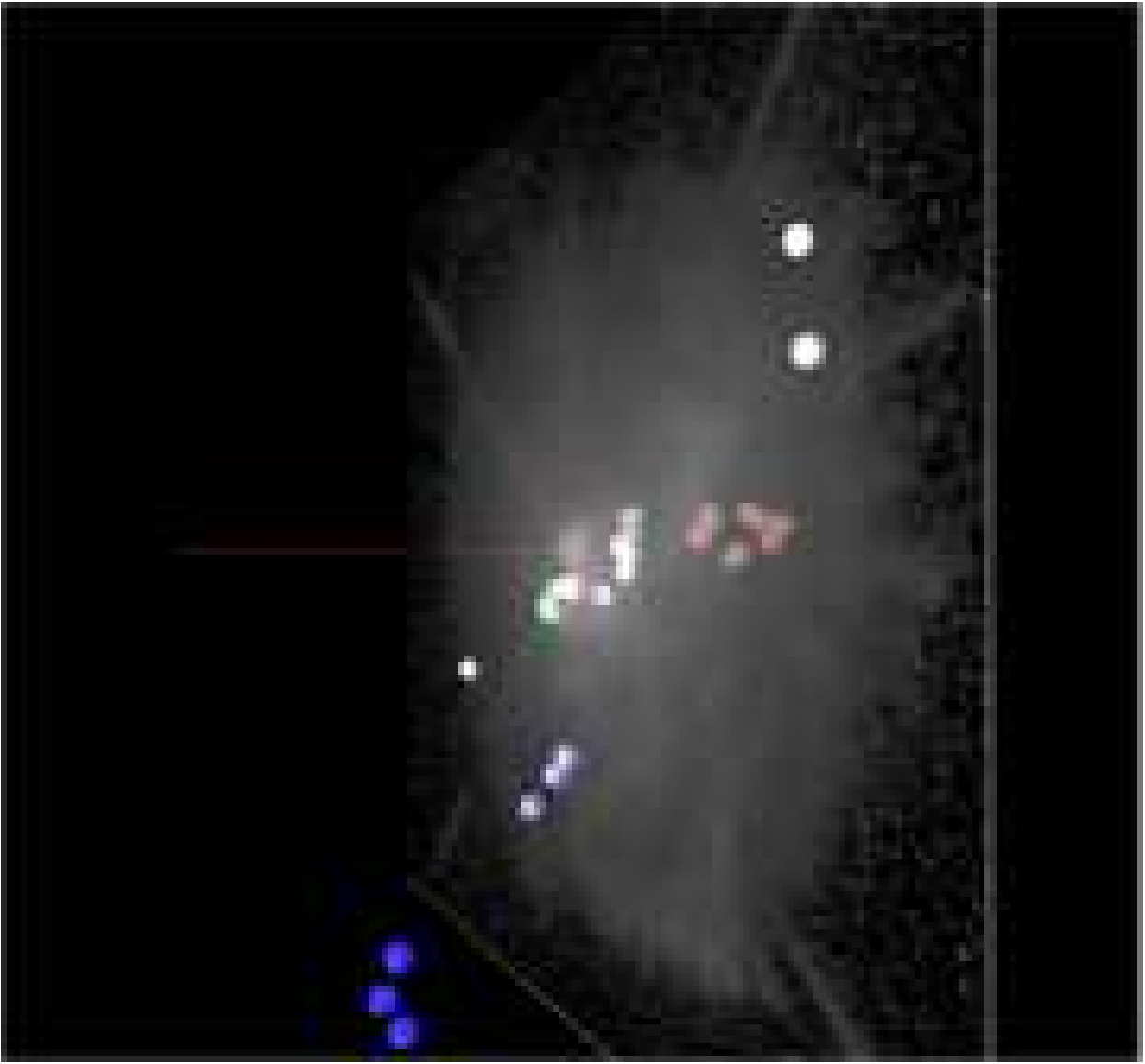}
\caption{The same as Figure 8, but rotated 60$\arcdeg$ west showing the positions 
relative to the plane of the sky.}  
\end{figure} 

\begin{figure}
\figurenum{10}
\epsscale{1.0}
\plotone{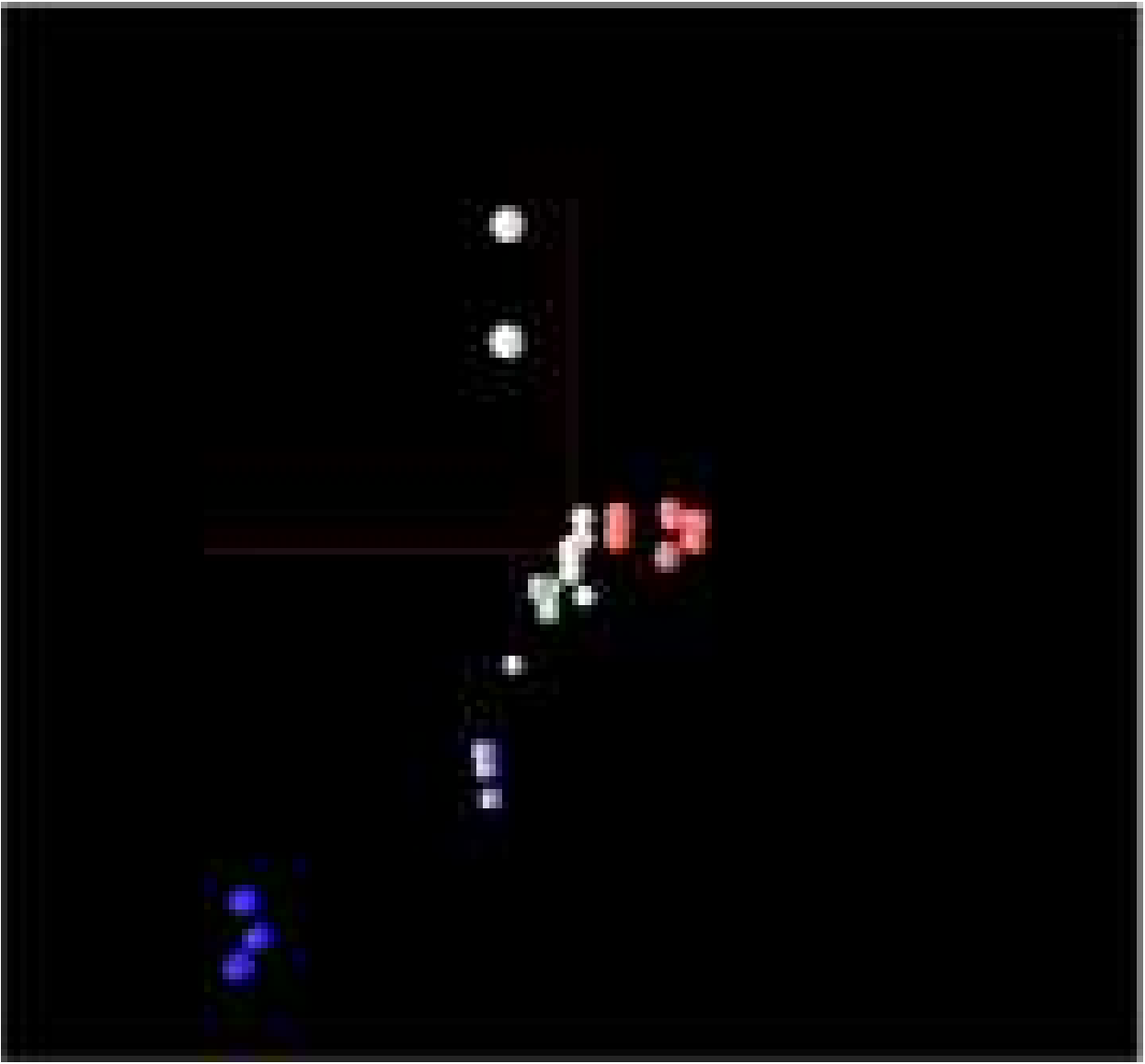}
\caption{The same as Figure 9, but rotated 90$\arcdeg$ west or edge-on.} 
\end{figure}

\input{tab1.tex}

\input{tab2.tex}

\input{tab3.tex}

\input{tab4.tex}

\input{tab5.tex}

\end{document}

%% file: tab1.tex

\begin{deluxetable}{lllll}
\tablecaption{Journal of New Observations}
\tablewidth{0pt}
\tablehead{
\colhead{Instrument}  &  \colhead{Date}  &  \colhead{Filter}  &
 \colhead{Exposure Times} & \colhead{Combined Images}    
}
\startdata
WFPC2   &   June 13, 2005  &  F410M  &   5$^{s}$, 16$^{s}$, 60$^{s}$ $\times$ 2 & F410{\it s}(21$^{s}$), F410{\it l}(120$^{s}$)\\ 
        &                  &  F547M  &   0.1$^{s}$, 0.5$^{s}$, 2$^{s}$, 5$^{s}$, 16$^{s}$ & F547{\it s}(0.6$^{s}$), F547{\it l}(21$^{s}$) \\
        &                  &  F656N  &   0.4$^{s}$, 2$^{s}$, 14$^{s}$, 60$^{s}$ & F656{\it s}(2.4$^{s}$), F656{\it l}(74$^{s}$)\\
        &                  &  F1042M & 0.1$^{s}$, 0.5$^{s}$, 3$^{s}$, 16$^{s}$ & F1042{\it s}(0.6$^{s}$), F1042{\it l}(19$^{s}$)  \\
ACS/HRC &   August 17, 2004 &  F550M/POL0V  &  0.2$^{s}$, 0.5$^{s}$, 5$^{s}$, 60$^{s}$ \\
        &                   &  F550M/POL60V &                 "                 \\
	&                   &  F550M/POL120V &                "                 \\
	&                   &  F658N/POL0V   &  1$^{s}$, 5$^{s}$, 40$^{s}$, 150$^{s}$ \\
	&                   &  F658N/POL60   &                "                       \\
	&                   &  F658N/POL120V &                "                       \\
\enddata
\end{deluxetable}

%% file: tab2.tex

\begin{deluxetable}{lllllllll} 
\tabletypesize{\scriptsize}
\rotate
\tablecaption{Measured Angular Shifts in Arc Seconds In the Different Filter Combinations} 
\tablewidth{0pt}
\tablehead{
\colhead{Feature ID} & \colhead{F410{\it s}} & \colhead{F410{\it l}} &  \colhead{F547{\it s}}  & \colhead{F547{\it l}} &
\colhead{F656{\it s}}  & \colhead{F656{\it l}}  & \colhead{F1042{\it s}} & \colhead{F1042{\it l}}
} 
\startdata
S knot A & 0.062 $\pm$ 0.004\tablenotemark{a} & 0.029 $\pm$ 0.004 &0.027 $\pm$ 0.002 & \nodata &  0.043 $\pm$ 0.005 & 0.037 $\pm$ 0.004 & 0.008 $\pm$ 0.007\tablenotemark{a} & 0.038 $\pm$ 0.005  \\
S knot B & \nodata  & \nodata & 0.041 $\pm$ 0.001 & 0.029 $\pm$ 0.004 & \nodata  & 0.037 $\pm$ 0.004 & 0.048 $\pm$ 0.003\tablenotemark{a} & 0.056 $\pm$ 0.003\tablenotemark{a} \\
S knot C & 0.052 $\pm$ 0.009\tablenotemark{a} & 0.015 $\pm$ 0.001 & 0.026 $\pm$ 0.004 & \nodata  & 0.019 $\pm$ 0.005 & 0.025 $\pm$ 0.001 & \nodata  & 0.065 $\pm$ 0.005\tablenotemark{a} \\ 
S knot D & 0.043 $\pm$ 0.005\tablenotemark{b} & 0.010 $\pm$ 0.003\tablenotemark{c} & \nodata  & \nodata  & 0.034 $\pm$ 0.004\tablenotemark{a} & \nodata  & \nodata  & \nodata  \\
S knot Y & \nodata & \nodata & \nodata & \nodata & \nodata & \nodata & \nodata & 0.048 $\pm$ 0.001 \\

SW knot A & 0.067 $\pm$ 0.015\tablenotemark{a} & 0.028 $\pm$ 0.002  & 0.012 $\pm$ 0.004 & \nodata & 0.034 $\pm$ 0.004 & \nodata & \nodata & \nodata \\
SW knot B & 0.025 $\pm$ 0.002 & 0.020 $\pm$ 0.003 & \nodata & \nodata & \nodata & \nodata & \nodata & \nodata \\
SW knot C & 0.020 $\pm$ 0.003 & 0.032 $\pm$ 0.003 & \nodata & \nodata & \nodata & \nodata & \nodata & \nodata \\ 
SW knot D & 0.043 $\pm$ 0.005 & 0.022 $\pm$ 0.002 & \nodata & \nodata & \nodata & \nodata & \nodata & \nodata \\ 
SW knot G & \nodata & 0.017 $\pm$ 0.003 & \nodata & \nodata & 0.050 $\pm$ 0.001 & \nodata & \nodata & \nodata \\
SW knot H & \nodata & \nodata & \nodata & \nodata & 0.035 $\pm$ 0.010 & \nodata & \nodata & \nodata \\
SW Clump & \nodata & \nodata & \nodata & \nodata & \nodata & \nodata & 0.008 $\pm$ 0.002 & 0.025 $\pm$ 0.003 \\

W arc A & 0.062 $\pm$ 0.010\tablenotemark{a} & 0.016 $\pm$ 0.006\tablenotemark{a} & 0.029 $\pm$ 0.003 & \nodata & \nodata & 0.025 $\pm$ 0.003  & \nodata & \nodata \\
W arc B & 0.040 $\pm$ 0.001 & 0.025 $\pm$ 0.001\tablenotemark{a} & \nodata & \nodata & 0.046 $\pm$ 0.003 & \nodata & \nodata & \nodata \\
W arc C & 0.050 $\pm$ 0.004\tablenotemark{d} & 0.042 $\pm$ 0.010\tablenotemark{e} & \nodata & \nodata & \nodata & \nodata & \nodata & 0.020 $\pm$ 0.006\tablenotemark{a} \\
W arc D & 0.059 $\pm$ 0.005\tablenotemark{f} & 0.030 $\pm$ 0.001\tablenotemark{g} & \nodata & \nodata & \nodata & \nodata & \nodata & \nodata \\

Inner W arc & \nodata & \nodata & \nodata & 0.023 $\pm$ 0.007 & \nodata & \nodata & \nodata & \nodata \\

NW arc A & \nodata & \nodata &  0.047 $\pm$ 0.004\tablenotemark{h} & 0.044 $\pm$ 0.004\tablenotemark{i} & \nodata & \nodata & \nodata & \nodata \\
NW arc B & \nodata & \nodata &  0.059 $\pm$ 0.003\tablenotemark{j} & 0.039 $\pm$ 0.013\tablenotemark{k} & \nodata & 0.023 $\pm$ 0.006\tablenotemark{l} &  \nodata & \nodata \\
NW arc C & \nodata & \nodata & 0.063 $\pm$ 0.002 & \nodata & \nodata & 0.031 $\pm$ 0.002 & \nodata & \nodata \\
NW arc D & \nodata & \nodata & 0.100 $\pm$ 0.004\tablenotemark{a} & 0.023 $\pm$ 0.005\tablenotemark{m} & \nodata  & 0.020 $\pm$ 0.003\tablenotemark{n} &  \nodata & \nodata \\
NW arc X & \nodata & \nodata & \nodata & \nodata & \nodata & \nodata & \nodata & 0.028 $\pm$ 0.004 \\
NW arc Y & \nodata & \nodata & \nodata & \nodata & \nodata & \nodata & \nodata & 0.036 $\pm$ 0.014 \\
NW arc Z & \nodata & \nodata & \nodata & \nodata & \nodata & \nodata & \nodata & 0.010 $\pm$ 0.010\tablenotemark{a} \\

NW knot & \nodata & \nodata & \nodata & \nodata & \nodata & \nodata & 0.032 $\pm$ 0.007 & 0.008 $\pm$ 0.003 \\

SE loop A & \nodata & \nodata & \nodata & 0.039 $\pm$ 0.008\tablenotemark{o} & \nodata & 0.074 $\pm$ 0.011\tablenotemark{p}  & \nodata & \nodata \\
SE loop B & \nodata & \nodata & \nodata & 0.053 $\pm$ 0.004 & \nodata & \nodata & \nodata & \nodata \\
SE knot & \nodata & \nodata & \nodata & 0.060 $\pm$ 0.006 & \nodata & 0.054 $\pm$ 0.005  & \nodata & \nodata \\

Inner S arc A & \nodata & \nodata & \nodata & 0.036 $\pm$ 0.014 & \nodata & 0.018 $\pm$ 0.005 & \nodata & \nodata \\
Inner S arc B & \nodata & \nodata & 0.040 $\pm$ 0.001 & \nodata & \nodata & \nodata & \nodata & \nodata \\
S arc A & \nodata & \nodata & \nodata & 0.040 $\pm$ 0.015 & \nodata & 0.034 $\pm$ 0.002  & \nodata & \nodata \\ 
S arc B & \nodata & \nodata & \nodata & 0.038 $\pm$ 0.011 & \nodata & 0.044 $\pm$ 0.002 & \nodata & \nodata \\
SW arc A & \nodata & \nodata & \nodata & 0.036 $\pm$ 0.011 & \nodata & \nodata & \nodata & \nodata \\
SW arc B & \nodata & \nodata & \nodata & 0.042 $\pm$ 0.002 & \nodata & 0.015 $\pm$ 0.005 & \nodata & \nodata \\
SW arc C & \nodata & \nodata & \nodata & \nodata & \nodata & 0.011 $\pm$ 0.004 & \nodata & \nodata \\
Finger & \nodata & \nodata & \nodata &  0.053 $\pm$ 0.009 & \nodata & 0.031 $\pm$ 0.004 & \nodata & \nodata \\

Arc 2 A & \nodata & \nodata & \nodata & 0.059 $\pm$ 0.002 & \nodata & 0.036 $\pm$ 0.003 & \nodata & \nodata \\
Arc 2 B & \nodata & \nodata & \nodata & 0.059 $\pm$ 0.003 & \nodata & 0.034 $\pm$ 0.008 & \nodata & \nodata \\
Arc 2 C & \nodata & \nodata & \nodata & 0.039 $\pm$ 0.005 & \nodata & \nodata & \nodata & \nodata \\
Arc 2 X  & \nodata & \nodata & \nodata & \nodata & \nodata & \nodata & \nodata & 0.048 $\pm$ 0.001 \\
Arc 2 Y   & \nodata & \nodata & \nodata & \nodata & \nodata & \nodata & \nodata & 0.071 $\pm$ 0.005 \\
Arc 2 Z  & \nodata & \nodata & \nodata & \nodata & \nodata & \nodata & \nodata & 0.075 $\pm$ 0.022 \\

Arc 1 A & \nodata & \nodata & \nodata & 0.053 $\pm$ 0.006 & \nodata & 0.037 $\pm$ 0.003 & \nodata & \nodata \\
Arc 1 B & \nodata & \nodata & \nodata & 0.050 $\pm$ 0.001\tablenotemark{q} & \nodata & 0.069 $\pm$ 0.003\tablenotemark{r}  & \nodata & \nodata \\
Arc 1 C & \nodata & \nodata & \nodata & 0.047 $\pm$ 0.007 & \nodata & \nodata & \nodata & \nodata \\
Arc 1 D & \nodata & \nodata & \nodata & 0.053 $\pm$ 0.005 & \nodata & 0.066 $\pm$ 0.010  & \nodata & \nodata \\
Arc 1 E & \nodata & \nodata & \nodata & 0.058 $\pm$ 0.002 & \nodata & \nodata & \nodata & \nodata \\
Arc 1 F &  \nodata & \nodata & \nodata & 0.042 $\pm$ 0.005 & \nodata & 0.062 $\pm$ 0.001 & \nodata & \nodata \\
Arc 1 G &  \nodata & \nodata & \nodata & 0.022 $\pm$ 0.004\tablenotemark{s} & \nodata & 0.073 $\pm$ 0.016\tablenotemark{t}  & \nodata & \nodata \\
Arc 1 H & \nodata & \nodata & \nodata & 0.006 $\pm$ 0.006\tablenotemark{a}  & \nodata & \nodata & \nodata & \nodata \\
Arc 1 I & \nodata & \nodata & \nodata & 0.056 $\pm$ 0.004 & \nodata & \nodata & \nodata & \nodata \\
Arc 1 J & \nodata & \nodata & \nodata & 0.019 $\pm$ 0.003 & \nodata & \nodata & \nodata & \nodata \\

Spike 1 A & \nodata & \nodata & \nodata & 0.025 $\pm$ 0.006 & \nodata & \nodata & \nodata & \nodata \\
Spike 1 B & \nodata & \nodata & \nodata & 0.086 $\pm$ 0.005 & \nodata & \nodata & \nodata & \nodata \\
Spike 3 A & \nodata & \nodata & \nodata & 0.035 $\pm$ 0.005 & \nodata & \nodata & \nodata & \nodata \\
Spike 4 A & \nodata & \nodata & \nodata & 0.064 $\pm$ 0.009  & \nodata & \nodata & \nodata & \nodata \\
Spike 4 B & \nodata & \nodata & \nodata & 0.024 $\pm$ 0.010 & \nodata & \nodata & \nodata & \nodata \\
\enddata
\tablenotetext{a}{Discrepant positions, not used.}
\tablenotetext{b}{ S knot D1}
\tablenotetext{c}{ S knot D2} 
\tablenotetext{d}{ W arc C1} 
\tablenotetext{e}{ W arc C2}
\tablenotetext{f}{ W arc D1}
\tablenotetext{g}{ W arc D2}
\tablenotetext{h}{ NW arc A1}
\tablenotetext{i}{ NW arc A2}
\tablenotetext{j}{ NW arc B1}
\tablenotetext{k}{ NW arc B2}
\tablenotetext{l}{ NW arc B3}
\tablenotetext{m}{ NW arc D1}
\tablenotetext{n}{ NW arc D2}
\tablenotetext{o}{ SE loop A1}
\tablenotetext{p}{ SE loop A2}
\tablenotetext{q}{ Arc 1 B1}
\tablenotetext{r}{ Arc 1 B2}
\tablenotetext{s}{ Arc 1 G1}
\tablenotetext{t}{ Arc 1 G2}
\end{deluxetable}

%% file: tab3.tex

\begin{deluxetable}{llllll}
\tabletypesize{\scriptsize}
\tablecaption{The Transverse Velocities, Direction of Motion and Positions} 
\tablewidth{0pt}
\tablehead{
 & \colhead{Radial Distance} & \colhead{Position Angle} &  \colhead{Weighted Mean}  & \colhead{Direction of}\\ 
 & \colhead{from Star}  & \colhead{from Star}  & \colhead{Transverse Velocity} & \colhead{Motion ($\phi$)}\\  
\colhead{Feature Id}  & \colhead{(arcsec)} & \colhead{(deg)}  &  \colhead{ V$_{T}$ (km s$^{-1}$)} & \colhead{(deg)}
} 
\startdata 
S knot A  & 0.97    & -176    & 36.7 $\pm$ 2.4  &  177 $\pm$ 5 \\
S knot B  & 0.85    & 176    & 37.7 $\pm$ 2.5  & 148 $\pm$ 7 \\
S knot C  & 0.63    & -163    & 22.8 $\pm$ 1.0  & 151 $\pm$ 12 \\
S knot D1  & 0.73   & -140    & 49.4 $\pm$ 6.9 & 46 $\pm$ 4\\
S knot D2  & 0.85   & -131    & 11.4 $\pm$ 4.3 & 21 $\pm$ 17\\
S knot Y   & 0.98   & 179    & 55   $\pm$ 2   & 147 $\pm$ 2\\

SW knot A  & 0.99   & -121    & 30.2 $\pm$ 2.5 & -101 $\pm$ 23\\
SW knot B  & 1.01   & -120    & 27.3 $\pm$ 2.6 & -3 $\pm$ 9\\
SW knot C  & 1.18   & -121    & 29.9 $\pm$ 3.1 & -95 $\pm$ 30\\
SW knot D  & 0.83   & -100    & 29.4 $\pm$ 3.3 & -63 $\pm$ 11\\
SW knot G1 & 1.23   & -115    & 19.2 $\pm$ 4.5 & -50 $\pm$ 12 \\
SW knot G2 & 1.23   & -115    & 57.6 $\pm$ 1.3 & -119 $\pm$ 4 \\
SW knot H  & 0.84   & -99    & 40.6 $\pm$ 15.1 & -104 $\pm$ 8 \\
SW Clump   & 0.97   & -135    & 17.7 $\pm$ 2.5  & -162 $\pm$ 15 \\

W arc A    & 1.71   & -92    & 30.5 $\pm$ 2.7 & -72 $\pm$ 30 \\
W arc B    & 1.69   & -100    & 48.4 $\pm$ 2.0 & -24 $\pm$ 23 \\
W arc C1    & 1.56  & -87    & 57.0 $\pm$ 5.0 & -57 $\pm$ 3 \\
W arc C2    & 1.70   & -91    & 48.2 $\pm$ 13.5 & -57 $\pm$ 5 \\
W arc D1    & 1.50   & -82    & 67.5 $\pm$ 7.8  & -69 $\pm$ 2 \\
W arc D2    & 1.64   & -72    & 34.1 $\pm$ 2.5  & -40 $\pm$ 2 \\
Inner W arc & 1.69   & -90    & 26.3 $\pm$ 9.5  & -139 $\pm$ 3 \\

NW arc A1   & 3.14   & -82    & 53.9 $\pm$ 6.0 & -122 $\pm$ 2 \\
NW arc A2   & 2.92   & -79    & 50.1 $\pm$ 5.2 & -145 $\pm$ 1 \\
NW arc B1   & 2.85   & -82    & 68.8 $\pm$ 3.5 & -87 $\pm$ 2 \\
NW arc B2   & 2.66   & -85    & 44.8 $\pm$ 19  & -120 $\pm$ 23 \\
NW arc B3   & 2.91   & -80    & 26.8 $\pm$ 8.3 & -27 $\pm$ 8 \\
NW arc C    & 2.63   & -86    & 52.0 $\pm$ 1.8 & -88 $\pm$ 19 \\
NW arc D1   & 2.57   & -92    & 26.8 $\pm$ 7.4 & -156 $\pm$ 11 \\
NW arc D2   & 2.69   & -90    & 23.5 $\pm$ 4.9 & -21 $\pm$ 4 \\
NW arc X    & 3.04   & -87    & 32.5 $\pm$ 5.8 & -107 $\pm$ 7 \\
NW arc Y    & 3.26   & -83    & 41.7 $\pm$ 16 &  -110 $\pm$ 8 \\ 
NW knot     & 0.46   & -58    & 14.6 $\pm$ 4.2 & -78 $\pm$ 5 \\

SE loop A1  & 2.68   & 137    & 44.9 $\pm$ 12 & 105 $\pm$ 12 \\
SE loop A2  & 2.72   & 140    & 85.2 $\pm$ 15.6 & 133 $\pm$ 7 \\
SE loop B   & 2.67   & 140    & 60.7 $\pm$ 6.4 &  161 $\pm$ 9 \\
SE knot     & 1.33   & 141    & 64.5 $\pm$ 5.9 &  124 $\pm$ 11\\
Inner S arc  A & 1.92   & -155    & 23.6 $\pm$ 6.8 &  165 $\pm$  8 \\
Inner S arc B & 1.59    & -162    & 48.1 $\pm$ 2.6 &  130 $\pm$ 3 \\
S arc A     & 2.82   & -152    & 50.0 $\pm$ 2.8  &  -163 $\pm$ 11 \\
S arc B     & 2.63   & -159    & 39.1 $\pm$ 2.5  &  152 $\pm$ 11  \\ 
SW arc A    & 1.80   & -138    & 41.6 $\pm$ 9.7    &  -175 $\pm$ 3 \\
SW arc B    & 1.97   & -129    & 40.0 $\pm$ 3.3  &  -151 $\pm$ 13  \\
SW arc C    & 1.82   & -139    & 13.2 $\pm$ 6.1    &  -66 $\pm$ 11 \\
Finger      & 2.56   & -117    & 39.6 $\pm$ 4.9    &  -150 $\pm$ 4 \\

Arc 2 A     & 3.59   & -151    & 63.1 $\pm$ 2.0    &  155 $\pm$ 7.5 \\
Arc 2 B     & 3.50   & -144    & 64.2 $\pm$ 3.7    &  -170 $\pm$ 15 \\
Arc 2 C     & 3.41   & -135    & 44.6 $\pm$ 6.8    &  -146 $\pm$ 5 \\
Arc 2 X     & 3.91   & -165    & 55.9 $\pm$ 2.0    &  -135 $\pm$ 2  \\
Arc 2 Y     & 4.09   & -175    & 81.5 $\pm$ 6.5    &  -171 $\pm$ 8 \\
Arc 2 Z     & 3.92   & -179    & 85.9 $\pm$ 26     &  174 $\pm$ 5\\

Arc 1 A     & 5.89   & -150    & 52.7 $\pm$ 3.0    &  -138 $\pm$ 5 \\
Arc 1 B1    & 6.28   & -147    & 56.9 $\pm$ 2.1    &  126 $\pm$ 6  \\
Arc 1 B2    & 6.35   & -147    & 88.1 $\pm$ 15.2   &  -109 $\pm$ 1 \\
Arc 1 C     & 6.31   & -145    & 54.4 $\pm$ 10.0   & 180 $\pm$ 2 \\
Arc 1 D     & 5.61   & -142    & 64.3 $\pm$ 6.8    & -175 $\pm$ 4 \\
Arc 1 E     & 4.92   & -161    & 67.1 $\pm$ 3.0    & 161 $\pm$ 4 \\
Arc 1 F     & 5.33   & -134    & 69.8 $\pm$ 1.8    & -113 $\pm$ 25 \\
Arc 1 G1    & 4.85   & -132    & 25.6 $\pm$ 6.2    & -177 $\pm$ 20 \\
Arc 1 G2    & 4.75   & -134    & 84.1 $\pm$ 22.4   & -121 $\pm$ 2 \\
Arc 1 I     & 5.17   & -111    & 64.4 $\pm$ 5.6    & 128  $\pm$ 5 \\
Arc 1 J     & 5.21   & -115    & 22.3 $\pm$ 5.4    & -158 $\pm$ 16 \\

Spike 1A    & 5.72   & -92    & 28.9 $\pm$ 8.1    & -138 $\pm$ 11 \\
Spike 1B    & 6.61   & -91    & 98.2 $\pm$ 6.9    & -92  $\pm$ 3 \\
Spike 3     & 6.40   & -71    & 40.5 $\pm$ 6.9    & -76  $\pm$ 11 \\
Spike 4A    & 6.05   & -65    & 73.5 $\pm$ 13.1   & -75  $\pm$ 4 \\
Spike 4B    & 6.76   & -61    & 27.8 $\pm$ 14.2   & -75  $\pm$ 6 \\
\enddata
\end{deluxetable}

%% file: tab4.tex

\begin{deluxetable}{llllllll}
\tabletypesize{\scriptsize}
\tablecaption{Summary of the Motions in the Circumstellar Ejecta}
\tablewidth{0pt}
\tablehead{
\colhead{Feature Id} & \colhead{V$_{x}$} & \colhead{V$_{y}$} & \colhead{$\phi$} & \colhead{V$_{z}$} & \colhead{$\theta$} & \colhead{V$_{Tot}$}  & \colhead{Comment} \\
&  \colhead{(km s$^{-1}$)}  &  \colhead{(km s$^{-1}$)}  & \colhead{(deg)} & \colhead{(km s$^{-1}$)} & \colhead{(deg)} & \colhead{(km s$^{-1}$)} &   
}
\startdata
S knot A  &  1.9  & -36.6  &  177 &  -17  &  -24.8  &  40.4  &  Slit V ap 3 \\
S knot B  &  20   &  -32  &   148 &  -17  &  -24.3  &  41.3  &  Slit V ap 3 \\
S knot C  &  11   &  -19.9 &  151  & -15  &  -33.3  &  27.3  &  between III Ap 1 and 2\tablenotemark{a}\\
S knot D1 &  35.5 & 34.3  &   46   &  -14  &  -15.8 &  51.3  &  Slit III ap 2 \\
S knot D2  &  4.1 &  10.6 &  21    &  -14 &  -50.8  &  18.0  &  Slit III ap 2\\
S knot Y   & 30  & -46  &   147   &  -16   & -16    &  57.3  &  Slit V ap 3 \\
SW knot A  & -29.6 & -5.8  & -101 & -14    &  -24.9 &  33.3  &  Slit III ap 2 \\
SW knot B  &  -1.4  & 27.3  & -3  & -14    & -27.1  &  30.7  &  Slit III ap 2 \\
SW knot C  &  -29.8  & -0.1 & -95 &  -14  &  -25.1  &  33.0  &  Slit III ap 2 \\
SW knot G1 &  -14.7  & 12.3  & -50 & -14 & -36 &  23.8  & Slit III ap 2 \\
SW knot G2 &  -50.4  & -27.9 & -119 & -14 & -13.7 & 59.3 & Slit III ap 2\\
SW Clump   &  -5.5  &  -16.8 & -162 & 2.5:  & 8: & 17.8  & Slits III ap 2 and II ap 3 \\
W arc A    &  -29   &  9.4   &  -72 &  0    & 0  & 30.5 & Slit III ap 3 \\
W arc B    &  -19.7 & 44.2   & -24  &  0    & 0  & 30.5 & Slit III ap 3 \\
W arc C1   &  -47.8 & 31     & -57  &  0    & 0  & 57   & Slit III ap 3 \\
W arc C2   &  -40.4 & 26.2   & -57  &  0    & 0  & 48.2 & Slit III ap 3 \\
W arc D1   &  -63   & 24.2   & -69  &  4    & 9.7 & 67.6 & Slit I ap 3 \\
W arc D2   &  -21.8 & 26.0   & -40  &  4    & 6.7 & 34.2  & Slit I ap 3 \\
NW arc A1   &  -45.7 & -28.6 & -122 & 19    & 19.4 & 57.1 & Slit III ap 4 \\
NW arc A2   &  -28.7 & -41.0 & -145 & 19    & 20.8 & 53.6 & Slit III ap 4 \\
NW arc B1   &  -68.7 & 3.6   & -87  & 9     & 7.5  & 69   & between III ap 3 and 4 \tablenotemark{b} \\
NW arc B2   &  -38.8 & -22.4 & -120 &  9    & 11.3 & 45.7 & between III ap 3 and 4 \tablenotemark{b} \\
NW arc B3   &  -12.2 & 23.9  & -27  &  9    & 18.6 & 28.3 & between III ap 3 and 4 \tablenotemark{b} \\
NW arc C    &  -51.9 & 1.8   & -88  &  9    &  9.8 & 52.8 & between III ap 3 and 4 \tablenotemark{b} \\
NW arc D1   &  -10.9 & -24.5 & -156 & 18.7  &  35  & 32.7 & Slit II ap 5 \\
NW arc D2   &  -8.4  & 21.9  & -21  & 18.7  &  38.5 & 30.0 & Slit II ap 5\\
NW arc X    &  -31.1  & -9.5 & -107 &  23   &  35.3 & 39.8 & between II ap 5 and 6 \tablenotemark{c} \\
NW arc Y    &  -39.2  & -14.2 & -110 & 23   &  28.9 & 47.6 & between II ap 5 and 6 \tablenotemark{c} \\
NW knot\tablenotemark{d}     &   -14.3  & 3.0 &  -78  & \nodata & \nodata & \nodata &  \\ 
SE Loop B &    19.8   & -57.4  & 161  & -23.5 & -21 &  65.1 & Slit III, new position\\
Inner S arc A & 6.1  &  -22.8 & 165 & -13 & -28.8 & 26.9 & Slit V ap 4 \\
S arc B       &  18.4 & -34.5 & 152 & -16 & -22.2 & 42.2 & between V ap 4 and 5 \tablenotemark{e} \\
Arc 2 A        & 26.7 & -57.2 & 155 & -19 & -16.7 & 65.9 & Slit V ap 5 \\
Arc 2 B       &  -11.1 & -63.2 & -170 & -19 & -16.5 & 67 & Slit V ap 5 \\
Arc 2 X      & -39.5  & -39.5 & -135  & -19 & -18.8 & 59.0 & Slit V ap 5 \\
Arc 1 A      & -35.2  & -39.1 & -138 & -37 & -35 & 64.4 & Slit V ap 7 \\
Arc 1 C      & -54.5 & -37    & 180  & -37 & -34 & 65.9 & Slit V ap 7\\
Arc 1 D      & -5.6  & -64.0  & -175 & -37 & -29.9 & 74.2 & Slit V ap 7\\
Spike 3      & -39.3 & 9.8    & -76  & -6 or 0 & -8.4 or 0 & 40.5 & Slit II ap 8 \tablenotemark{f}\\
Spike 4B     & -26.8 & 7.2    & -75  & -6 or 0  & -12 or 0 & 28.4 & slit I ap 6 \\ 
  \enddata
  \tablenotetext{a}{Velocity interpolated between apertures 1 and 2.}     
  \tablenotetext{b}{Velocity interpolated between apertures 3 and 4.}
 \tablenotetext{c}{Velocity interpolated between apertures 5 and 6.}   
 \tablenotetext{d}{Same aperture as star.}
\tablenotetext{e}{Velocity interpolated between apertures 4 and 5}
\tablenotetext{f}{Two velocities are measured at 35 and 41 km s$^{-1}$. It is not possible to tell which is the appropriate velocity for this feature, but it makes little difference for the result.}

  \end{deluxetable}

%% file: tab5.tex

\begin{deluxetable}{lllll}
\tablecaption{Summary of Vector Motions and Ejection Ages for the Major Features}
\tablewidth{0pt}
\tablehead{
\colhead{Feature} & \colhead{V$_{Tot}$} & \colhead{$\theta$} & \colhead{Mean $\phi$\tablenotemark{a}}  & \colhead{Ejection Age}\\
& \colhead{(km s$^{-1}$)}  & \colhead{(deg)} & \colhead{(deg)}   &  \colhead{(years)} 
}
\startdata
NW Arc  &   45.7 $\pm$  4  & 22 $\pm$ 7 &  -98 $\pm$ 13\tablenotemark{b}  &  500 $\pm$ 50\\
Arc 1   &   68.2 $\pm$ 2.5 & -33 $\pm$ 3 & -161 $\pm$ 13\tablenotemark{c}  & 800 $\pm$ 50 \\
Arc 2   &   64 $\pm$ 2.1   & -17 $\pm$ 1 &  -174 $\pm$ 8\tablenotemark{d}     & 400 $\pm$ 15\\
W Arc   &   43.7 $\pm$ 4.8   & $\sim$ 0 $\pm$ 3 & -53 $\pm$ 7  & 300 $\pm$ 30 \\
SW Knots & 36 $\pm$ 5.5    & -25 $\pm$ 3 & -86 $\pm$ 14\tablenotemark{e}  & 250 $\pm$ 50 \\ 
S  Knots & 41.6 $\pm$ 5    & -27 $\pm$ 4 & 156 $\pm$ 6\tablenotemark{f}  & 157 $\pm$ 25\\
S Arc\tablenotemark{g}    & 42.2 $\pm$ 2.5  & -22         &  174 $\pm$ 8   & 480 $\pm$ 25\\
SE Loop  & 65.1 $\pm$ 4.6  & -21 & 133 $\pm$ 13      & 320 $\pm$ 20\\
SW Clump & $\sim$ 18 & 8:  & -162  & 500 \\
``spikes'' & 30 -- 40  &   $\sim$ 0 & \nodata  \tablenotemark{h} & 1300 -- 1700 \\
\enddata 
\tablenotetext{a}{Mean position angle of the tranverse motions (Table 3)}
\tablenotetext{b}{With a lack of measurements at the tip and in the hook of the NW Arc, this result is uncertain.}
\tablenotetext{c}{Arc1 excluding knots I and J.}
\tablenotetext{d}{Arc 2 excluding knot C.}
\tablenotetext{e}{Excluding knot B, but the SW Knots may be expanding away from each other.}
\tablenotetext{f}{S Knots excluding D1 and D2.}
\tablenotetext{g}{knot B only; an alternate choice of Doppler velocity gives a vector direction away from us at +16$\arcdeg$, but the same ejection age.}
\tablenotetext{h}{Spikes 3 and 4 have the same $\phi$ of -75$\arcdeg$.} 

\end{deluxetable}

%% file: ms.bbl
\begin{thebibliography}{}
\bibitem[Barvainis et al 1987]{Bar87}Barvainis, R., McIntosh, G. \& Predmore, C. R. 1987, Nature, 329, 613
\bibitem[Bowers, Johnston, \& Spencer 1983]{Bow83}Bowers, P. F., Johnston, K. J., \& Spencer, J. H. 1983, \apj, 274, 733
\bibitem[Danchi et al. 1994]{Dan94}Danchi, W.C., Bester, M., Degiacomi, C.G., Greenhill, L.J., \& Townes, C.H.  1994, \aj, 107, 1469
\bibitem[de Jager 1998]{deJ98}de Jager, C. 1998, \aapr, 8, 145
\bibitem[Flower 1977]{PJF77}Flower, P. J. 1977, \aap, 54, 31  
\bibitem[Gilliland \& Dupree 1996]{GD96} Gilliland, R.L. \& Dupree, A. K. 1996, \apj, 463, L29
\bibitem[Herbig 1970]{GH70}Herbig, G.H. 1970a, Mem. Soc. Roy. Liege, 19, 13
\bibitem[Herbig 1972]{Her72}Herbig, G.H.  1972, \apj, 172, 375
\bibitem[Humphreys 1975]{RMH75}Humphreys, R. M. 1975, \pasp, 87, 433
\bibitem[Humphreys \& Davidson (1979)]{HD79}Humphreys, R.M. \& Davidson, K. 1979, \apj, 232,
 409
\bibitem[Humphreys \& Davidson 1994]{HD94}Humphreys, R.M. \& Davidson, K. 1994, \pasp, 106, 1025 
\bibitem[Humphreys et al 2005]{RMH05}Humphreys, R. M., Davidson, K., Ruch, G., \& Wallerstein, G. 2005, \aj, 129, 492
\bibitem[Humphreys \& McElroy 1984]{HM84}Humphreys, R. M. \& McElroy, D. B. 1984, \apj, 284,
 565
\bibitem[Kastner \& Weintraub 1998]{KW98}Kastner, J. H. \& Weintraub, D.A. 1998, \aj, 115, 1592
\bibitem[Kemball \& Diamond 1997]{Kem97}Kemball, A. J. \& Diamond, P. J. 1997, \apjl, 481, L111
\bibitem[Kwok 1976]{Kwok76}Kwok, S. 1976, JRASC, 70, 49 
\bibitem[Lada \& Reid 1978]{Lad78}Lada, C.J. \& Reid, M.J.  1978, \apj, 219, 95
\bibitem[Le Sidaner \& Le Betre 1996]{LL96}Le Sidaner, P. \& Le Betre, T. 1996, \aap, 314, 8
96
\bibitem[Levesque et al 2005]{LM05}Levesque, E.M., Massey, P., Olsen, K. A. G., Plez, B., Jo
sselin, E., Maeder, A., \& Meynet, G. 2005, \apj, 628, 973  
\bibitem[Marvel 1997]{Mar97}Marvel, K.B.  1997, \pasp, 109, 1286
\bibitem[Masheder et al 1999]{Mash99}Masheder, M. R. W. et al 1999, New Astronomy, 43, 563 
\bibitem[Massey et al 2006]{PM06}Massey, P., Levesque, E. M., \& Plez, B. 2006, \apj, 646, 1
203
\bibitem[Monnier et al 1999]{Mon99}Monnier, J. D., Tuthill, P. G., Lopez, B., Cruzalebes, P., Danchi, W. C. \& Haniff, C. A. 1999, \apj, 512, 351 
\bibitem[Monnier et al 2004]{Mon04}Monnier, J. D. et al. 2004, \apj, 605, 436 
\bibitem[Morris \& Bowers 1980]{MB80}Morris, M. \& Bowers, P. F., 1980, \aj, 85, 724
\bibitem[Muller et al 2006]{Mull06}Muller, S., Dinh-V-Trung, Lim, J. Hirano, N., Muthu, C, \& Kwok, S. 2006, \apj, submitted 
\bibitem[Richards, Yates \& Cohen 1998]{RYC98}Richards, A. M. S., Yates, J. A., \& Cohen, R. J. 1998, \mnras, 299, 319
\bibitem[Shinnaga et al 2004]{Shi04}Shinnaga, H., Moran, J. M., Young, K. H. \& Ho, P. T. P. 2004, \apjl, 616, L47
\bibitem[Schuster et al 2006]{Sch06}Schuster, M. T., Humphreys, R. M., \& Marengo, M 2006, \aj, 131, 603 
\bibitem[Schwarzschild 1975]{MS75} Schwarzschild, M. 1975, \apj, 195, 137 
\bibitem[Smith et al. 2001]{Smi01}Smith, N., Humphreys, R. M., Davidson, K., Gehrz, R.  D., Schuster, M. T. \& Krautter, J.  2001, \aj, 121, 1111
\bibitem[Szymczak \& Cohen 1997]{Szy97}Szymczak, M. \& Cohen, R. J. 1997, \mnras, 288, 945
\bibitem[Tuthill, Haniff \& Baldwin 1997]{Tut97} Tuthill, P. G., Haniff, C.A. \& Baldwin, J. E. 1997, \mnras, 285, 529 
\bibitem[Van Blerkom \& Van Blerkom 1978]{VB78}Van Blerkom, J. \& Van Blerkom, D. 1978, \apj
, 225, 482
\bibitem[Vlemmings et al 2002]{Vlem02}Vlemmings, W. H. T., Diamond, P. J. \& van Langevelde, H. J. 2002, \aap, 394, 589
\bibitem[Vlemmings et al 2004]{Vlem04}Vlemmings, W. H. T., van Langevelde, H. J., \& Diamond, P. J. 2004, Mem S. A. It., 75, 282 
\bibitem[Wallerstein 1977]{GW77}Wallerstein, G. 1977, \apj, 211, 170
\bibitem[Wallerstein 1978]{Wall78}Wallerstein, G. 1978, Observatory, 98, 224
\bibitem[Wittkowski, Langer, \& Weigelt 1998]{Witt}Wittkowski, M., Langer, N., \& Weigelt, G. 1998, \aap, 340, L39  
\end{thebibliography}
